\documentclass[preprint]{aastex631}

\newcommand{\sgr}{\mbox{SGR\,J1935+2154~}}
\newcommand{\sgrnosp}{\mbox{SGR\,J1935+2154}}

\newcommand{\swiftnosp}{\mbox{\textit{Swift}}}

\newcommand{\fermi}{\mbox{\textit{Fermi}-GBM~}}
\newcommand{\ferminosp}{\mbox{\textit{Fermi}-GBM}}
\newcommand{\chandra}{\mbox{\textit{Chandra~}}}
\newcommand{\XMM}{\mbox{\textit{XMM-Newton~}}}

\newcommand{\ep}{$E_{\rm peak}$~}

\newcommand{\lkt}{$kT_{\rm Low}$~}
\newcommand{\lktnosp}{$kT_{\rm Low}$}
\newcommand{\hkt}{$kT_{\rm High}$~}
\newcommand{\hktnosp}{$kT_{\rm High}$}
\newcommand{\mkt}{$kT_{\rm M}$~}
\newcommand{\mktnosp}{$kT_{\rm M}$}

\font\fiverm=cmr5
\def\teq#1{$\, #1\,$}
\def\sigt{\sigma_{\hbox{\fiverm T}}}
\def\taut{\tau_{\hbox{\fiverm T}}}

\font\fiverm=cmr5

\usepackage{comment}
\usepackage{amsmath, soul}
\usepackage{multirow}

\graphicspath{{./}{figures/}}

\shorttitle{\sgrnosp}
\shortauthors{Keskin et al.}

\begin{document}

 \title{Concise Spectrotemporal Studies of Magnetar \sgr Bursts}

\correspondingauthor{\"Ozge Keskin}
\email{ozgekeskin@sabanciuniv.edu}

\author[0000-0001-9711-4343]{\"Ozge Keskin}
\affiliation{Sabanc\i~University, Faculty of Engineering and Natural Sciences, \.Istanbul 34956 Turkey}

\author[0000-0002-5274-6790]{Ersin G\"o\u{g}\"u\c{s}}
\affiliation{Sabanc\i~University, Faculty of Engineering and Natural Sciences, \.Istanbul 34956 Turkey}

\author[0000-0002-1861-5703]{Yuki Kaneko}
\affiliation{Sabanc\i~University, Faculty of Engineering and Natural Sciences, \.Istanbul 34956 Turkey}

\author[0009-0000-9126-7824]{Mustafa Demirer}
\affiliation{Sabanc\i~University, Faculty of Engineering and Natural Sciences, \.Istanbul 34956 Turkey}

\author[0000-0002-1688-8708]{Shotaro Yamasaki}
\affiliation{Department of Physics, National Chung Hsing University, 145 Xingda Rd., South Dist., Taichung 40227, Taiwan}

\author[0000-0003-4433-1365]{Matthew G. Baring}
\affiliation{Department of Physics and Astronomy - MS 108, Rice University, 6100 Main Street, Houston, Texas 77251-1892, USA}

\author[0000-0002-0633-5325]{Lin Lin}
\affiliation{Department of Astronomy, Beijing Normal University, Beijing 100875, China}

\author[0000-0002-7150-9061]{Oliver J. Roberts}
\affiliation{Science and Technology Institute, Universities Space and Research Association, 320 Sparkman Drive, Huntsville, AL 35805, USA.}

\author[0000-0003-1443-593X]{Chryssa Kouveliotou}
\affiliation{Department of Physics, The George Washington University, 725 21st Street NW, Washington, DC 20052, USA}
\affiliation{Astronomy, Physics, and Statistics Institute of Sciences (APSIS), The George Washington University, Washington, DC 20052, USA}


\begin{abstract}

\sgr has truly been the most prolific magnetar over the last decade: It has been entering into burst active episodes once every 1-2 years since its discovery in 2014, it emitted the first Galactic fast radio burst associated with an X-ray burst in 2020, and has emitted hundreds of energetic short bursts. Here, we present the time-resolved spectral analysis of 51 bright bursts from \sgrnosp. Unlike conventional time-resolved X-ray spectroscopic studies in the literature, we follow a two-step approach to probe true spectral evolution. For each burst, we first extract spectral information from overlapping time segments, fit them with three continuum models, and employ a machine learning based clustering algorithm to identify time segments that provide the largest spectral variations during each burst. We then extract spectra from those non-overlapping (clustered) time segments and fit them again with the three models: the cutoff power-law model, the sum of two blackbody functions, and the model considering the emission of a modified black body undergoing resonant cyclotron scattering, which is applied systematically at this scale for the first time. Our novel technique allowed us to establish the genuine spectral evolution of magnetar bursts. We discuss the implications of our results and compare their collective behavior with the average burst properties of other magnetars.

\end{abstract}



\section{Introduction} \label{sec:intro}

High-energy transient activity from magnetars falls into three general categories: short and intermediate bursts, and Giant Flares (GFs). The most frequently observed events are the short bursts, with durations ranging from a few milliseconds to a few seconds, peaking at $\sim$ 0.1 s; their energies approach $10^{41}$ ergs \citep{EG01}. GFs cover the other end of the magnetar burst properties with much harder spectra and longer durations. These are the most energetic events, releasing over $10^{44}$ ergs in several minutes \citep{Hurley99, Palmer05}. Apart from their longer durations, they also exhibit a unique morphology: a spectrally hard initial short spike, followed by a longer tail, that oscillates at the spin frequency of the parent neutron star. Finally, intermediate events exhibit energetics and durations in between short bursts and GFs. Their durations range from a few seconds to a few tens of seconds with energies extending up to $\sim10^{42}$ ergs \citep{Alaa01, CK01}.

The run-of-the-mill magnetar bursts are generally attributed to local or global yielding of the solid neutron star crust under the influence of enormous internal and external magnetic pressure, along with excitation of modes by instabilities in the magnetosphere \citep{TD95, TD01}. This idea was put into action by \citet{Lander23} who showed that the build-up of elastic stress in the crust results in its failure and in energy release. Another possible process that can generate magnetar bursts is magnetic field line reconnection \citep{TD95,lyu03}.

Hard X-ray spectral analyses of magnetar bursts are crucial towards a better understanding of the physical mechanisms responsible for these intriguing phenomena. Studies so far have shown that magnetar bursts can be modeled almost equally well with a thermal model as with the sum of two black bodies (BB+BB), or a non-thermal model, such as a power law with a high-energy exponential cutoff (COMPT) \citep{von12, Lin12}. This poses a puzzle in the understanding of the underlying mechanism of the observed bursts, as follows. According to the crustal fracturing scenario \citep{TD95}, a fireball by the plasma of trapped $e^-$$-e^+$ pairs and photons forms in the closed magnetic field lines by Alfven waves released from the crust following the cracking. Therefore, thermal radiation could be expected from such regions in quasi-equilibrium. On the other hand, the observed magnetar synchrotron-like non-thermal radiation spectra might be an indication of particle interactions via magnetic reconnection. Recently, \citet{Yamasaki20} studied the spectral modification of extremely energetic magnetar flares by resonant cyclotron scattering and showed that the scattering process may alter the emerging radiation from these events. In this model, photons are emitted via a mechanism that depends only on the temperature of the fireball near the neutron star surface interact with magnetospheric particles, resulting in significant changes in the emission spectrum.

Time-resolved X-ray spectroscopy of magnetar bursts is an important probe to reveal spectral evolution throughout their highly-complex emission episodes. One of the most comprehensive time-resolved burst spectral studies was performed by \citet{Israel08} on the SGR 1900+14 bursts observed with the Burst Alert Telescope (BAT) and X-ray Telescope (XRT) on board the \textit{Neil Gehrels Swift} Observatory (hereafter \swiftnosp) in March 2006. Although the BB+BB model was at the forefront of this study including over 700 extracted spectra, a significant number of the spectra were also fitted with non-thermal models. Detailed analyses based on the BB+BB model revealed significant spectral evolution, even during short magnetar bursts. \citet{Younes14} also performed time-resolved spectral analysis of the 63 brightest bursts from SGR\,J1550$-$5418 observed with the Gamma-ray Burst Monitor (GBM) on the \textit{Fermi} Gamma-ray Space Telescope (hereafter \textit{Fermi}) and demonstrated flux dependent variations between temperature obtained with the BB+BB model and the area of the emitting region. 

Time binning for the extraction of spectra in time-resolved spectroscopy of bursts, however, is quite arbitrary. In earlier studies, the subsequent spectra were usually obtained from time intervals determined based upon the signal-to-noise ratio of its light curve. In other words, the time segments are not determined by taking into account the spectral changes in advance, but the burst is divided blindly into time segments to contain a certain number of photons to ensure acceptable statistics for the spectral analysis. Under such circumstances, it would not be possible to elucidate the true spectral evolution of the observed burst emission. 

One possible way to overcome the problem of arbitrariness in time-resolved spectroscopy is to employ sequential spectra extracted from the shortest possible time intervals (in terms of counts) and allow these segments to overlap subsequently. This way, it would naturally be possible to reveal real spectral evolution, which in turn could help uncover more dominant underlying mechanism. This sliding time window approach is a methodology often used in timing analysis, e.g., searching for time-dependent burst oscillation behaviors \citep[see e.g.,][]{Stroh98}. In this study, we apply a similar approach for the first time in the spectral analysis of SGR bursts; 
namely, ``overlapping time-resolved spectroscopy", using the brightest bursts from \sgrnosp. We subsequently apply a machine learning based clustering algorithm to form non-overlapping time intervals with significant spectral variations and perform ``clustering-based" time-resolved spectral analysis of these bursts. Hence, the resulting time segments are expected to precisely reveal the burst spectral evolution. 

To perform spectral modeling of \sgr bursts, we employ commonly-used BB+BB and COMPT models. In addition, we also employ the model of a modified black body (MBB) spectrum \citep{Lyubarsky_2002} that undergoes resonant cyclotron scattering (RCS), the combination of which is applied systematically to the SGR burst spectral analysis at this scale for the first time. The MBB-RCS model was developed by \citet{Yamasaki20} to account for the magnetospheric effects on thermal emission in the context of magnetar flares. \citet{Lyubarsky_2002} had previously suggested extreme magnetic fields of magnetars could modify the emerging radiation from its surface, incurring significant deviations from a Planckian that result in a flat photon spectrum at low energies. However, the pure MBB model can underestimate the spectra at high energies by not taking the magnetospheric scatterings into account. The MBB-RCS model \citep{Yamasaki20} assumes that the photons emitted from the active burst region escape to infinity after just one resonant scattering by magnetospheric charges, thereby accounting for tails in the observed spectra at high energies. Testing the model with energetic flares of SGR 1900+14 resulted in a good agreement with the spectra of intermediate flares, but not with the giant flare observed in 1998.

\sgr was discovered on 2014 July 5, when a short burst triggered \swiftnosp/BAT. Follow-up observations with \chandra and \XMM revealed its spin period ($P \sim$ 3.24 s) and spin-down rate $(\dot{P} = 1.43 \times 10^{-11}$\,s\,s$^{-1})$, corresponding to a magnetic field strength of $B \sim 2.2 \times 10^{14}$ G, and thus confirming the magnetar nature of the source  \citep{Israel06}. Since its discovery, \sgr has been the most prolific transient magnetar ever observed: it is burst-active almost annually including multiple (in the range of thousand) short bursts \citep{Lin2016, Lin2020}, an intermediate flare on 2015 April 12 \citep{Kozlova16}, and a burst forest on 2020 April 27 \citep{YK21}. Just hours after the burst forest, \sgr emitted an X-ray burst \citep[e.g.,][]{mereghetti20} coincident with a Fast Radio Burst \citep[FRB;][]{chime20,bochenek}, which was the first Galactic detection of these events. This coincidence is the first indicator that magnetars residing in distant galaxies may also be the origin of FRBs \citep{petroff,petroff21}.

This paper is organized as follows: In Section \ref{sec:info}, we introduce instrument, data, the \sgr bursts that we studied, and their spectral data extraction process. In Section \ref{sec:spectro}, we explain the steps in the spectral analysis: first overlapping and then clustering-based time-resolved spectroscopy. The results are presented and discussed in Section \ref{sec:resndis}.


\section{Instrument \& \sgr Burst Observations} \label{sec:info}

\textit{Fermi} has been providing an enormous amount of data that allows for studying a wide range of gamma-ray transient events over the last 15 years. It carries two instruments: the Gamma-ray Burst Monitor (GBM; $\sim$ 8\,keV$-$40\,MeV) and the Large Area Telescope (LAT; $\sim$ 20\,MeV$-$300\,GeV). GBM consists of 12 sodium iodide (NaI) detectors ($\sim$ 8\,keV$-$1\,MeV) and two bismuth germanate (BGO) detectors ($\sim$ 200\,keV$-$40\,MeV) \citep{Meegan2009}. Since the bulk of emission from magnetar bursts is typically seen in $\lesssim$ 200\,keV, we only used the data of NaI detectors for this study. We employed Continuous Time-Tagged Event (CTTE) data of GBM, which provides the finest time (2 $\mu$s) and energy (128 channels) resolutions. Our investigations were performed in the 8$-$200\,keV energy band with 4 ms minimum time resolution. For each burst, we included data collected with the three brightest NaI detectors\footnote{Detectors with the lowest detector-to-source angle at the time of the event.} with the detector-to-source angle being less than $60^{\rm o}$. Additionally, we excluded detectors if they are partially or fully blocked by other parts of the spacecraft as obtained using the GBMBLOCK software provided by the GBM team. 

We selected 51 \sgr bursts observed with \fermi between 2014 and 2022 for our time-resolved spectral investigations, based on the results of time-integrated analyses \citep[][Lin et al. in preparation]{Lin2016, Lin2020}. In particular, we chose the bursts that contain at least 2400 background-subtracted counts to ensure that they have enough statistics for time-resolved spectral analysis (See Appendix \ref{appA}). Our investigations are done in two stages for each burst: First, we define the overlapping time segments with which we obtain spectral parameters and use a machine learning clustering algorithm to obtain ``change points" for the parameter evolution. Then, we analyze spectral data extracted from the (non-overlapping) time intervals between these change points to better characterize spectral evolution. 

\begin{figure}[htbp]
    \centering
    \epsscale{1.15}
    \includegraphics[width = 14 cm, trim=82 280 45 265, clip]{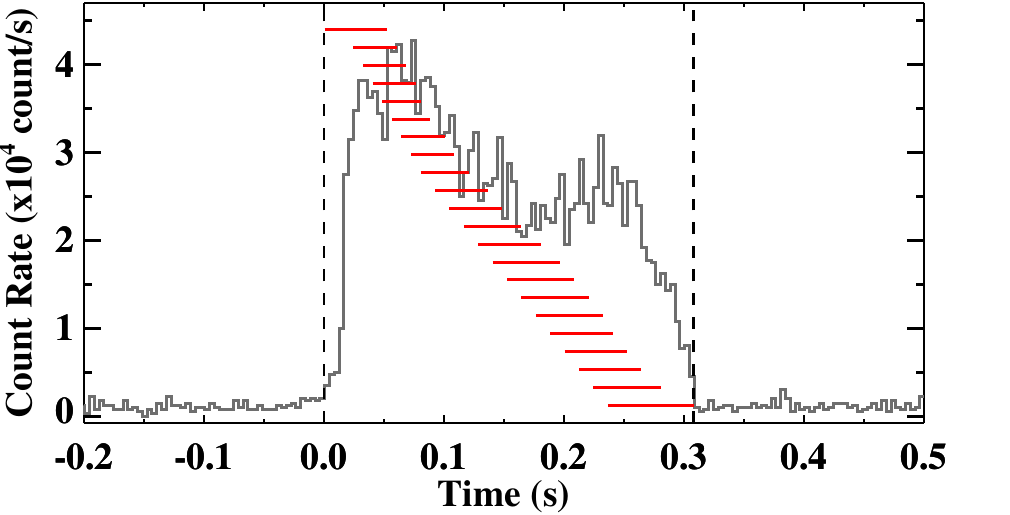}
       \caption{Light curve of an \sgr burst detected at 2021 January 29 10:35:39.918 UTC as seen with the brightest detector (n5). The vertical dashed lines show Bayesian Block duration start and end times, respectively. The red horizontal lines represent the 22 overlapping time segments with each subsequent segment having an overlap of 80\%. We note that the vertical values corresponding to the red lines are arbitrary, and chosen only for display purposes.
}
    \label{lc_ex}
\end{figure}

In the first stage of our time-resolved spectral studies, we required each time segment to overlap with the previous time segment by 80\%. In other words, the subsequent time segment does not start from the end of the previous one, but from the time that is one fifth of the previous segment (see Figure \ref{lc_ex} for an example). We also required a minimum of 1200 background-subtracted counts in each time segment (See Appendix \ref{appA}). This way, burst spectral parameters are expected to be well-constrained and statistically reliable throughout each burst. We note that for each burst, we calculated duration\footnote{The details of duration calculation as well as background estimation can be found in Appendix \ref{appA}.} via Bayesian Block method optimized for photon counting time series \citep{scargle2013}, and started the first time segment from the beginning of Bayesian Block duration.  


Time segments starting before the peak of a burst are prone to end at the peak although they are shifted by 20\% of the time length of their previous ones due to the fact that flux rise timescale in short magnetar bursts is usually shorter than that of decay \citep{EG01}. This yields an accumulation of time segments at the beginning of the bursts. In those cases, we avoided the accumulation of time segments by reducing the overlap by 5\% (75\%, 70\%, etc.) until the endpoint of the subsequent time segment ends later than the end of the previous one. In the end, we obtained in total 1343 overlapping time segments, hence spectra, from the 51 bursts for the first stage of our spectroscopy investigation.


\section{Spectral Analyses \& Results}
\label{sec:spectro}

We performed our spectral analysis with XSPEC (version 12.12.1) using Castor Statistics \citep[C-stat;][]{Castor}. We also generated Detector Response Matrices (DRM) with the GBM Response Generator\footnote{\url{https://fermi.gsfc.nasa.gov/ssc/data/analysis/rmfit/DOCUMENTATION.html}} released by \fermi team. As mentioned above, we fit the spectrum of each time segment (in 8$-$200\,keV) with three different photon models: an exponentially cutoff power law model (COMPT\footnote{$f(E) = A \exp{[-E(2+\Gamma)/E_{\rm peak}]}(E/50 {\rm keV})^{\Gamma}$}), the sum of two black body functions (BB+BB), and a modified black body with resonant cyclotron scattering (MBB-RCS\footnote{We generated a table model to be used in XSPEC by following the prescription of \citet{Yamasaki20}. The table model covers the energy range from 5 to 300 keV, and the parameter grid of T$_{\rm eff}$ consists of 79 values between 1 and 40 keV with increments of 0.5 keV. This table model can be downloaded from
\url{https://zenodo.org/records/10485159}.}). 

 Based on the C-stat value obtained with each model fit, we employed the Bayesian Information Criterion (BIC; \citealt{Schwarz1978, Liddle07}) to evaluate the improvement of one model compared to the others, as follows: \[\text{BIC} = \text{C-stat} + m \ln N \] where $m$ is the number of parameters in the photon model and $N$ is the number of data points. By comparing the BIC values of each of the three photon model fits in pairs, we determined a statistically preferred model that is with significantly lower BIC ($\Delta$BIC $>$ 10, which corresponds to the Bayes factor of $\sim$150, indicating the confidence level $>$99\% for the likelihood ratio; \citealt{Bayes95}). If the difference in BIC values is small, i.e., $\Delta$BIC $\leq$ 10, then both models are equally preferred. 

Following the fits, we also calculated the photon and energy flux of each time segment in the energy range of 8$-$200\,keV based on the fit parameters of each photon model. Note that all errors reported throughout the paper are at the confidence level of 1$\sigma$. We also require model parameters to be well-constrained with their 1$\sigma$ errors (i.e., model parameter $\pm$ 1$\sigma$ error must be viable values) for the fits to be considered acceptable.


\subsection{Overlapping Time-Resolved Spectroscopy} 

Out of the 1343 spectra modeled, we found that 1322 of them (i.e., 98.4\% of the sample) are described well with COMPT, meaning that they were deemed preferred based on the BIC values. The thermal models, on the other hand, perform nearly equally in fitting: BB+BB and MBB-RCS models are statistically preferred for 542 (40.4\%) and 551 (41\%) out of the 1343 time segments, respectively. Here, ``preferred" means that the BIC value of a model is either significantly lower than the other two models ($\Delta$BIC $>$ 10, hence the model is the only preferred model) or comparable to the other model(s) ($\Delta$BIC $\leq$ 10, hence two or all three models are comparably preferred).

As stated before, our aim is to perform a clustering-based time-resolved spectral analysis to clearly reveal spectral evolution throughout short magnetar bursts. Therefore, we identified the significant spectral change points of the bursts using a machine learning based clustering algorithm with the results of spectral analysis of overlapping time segments. To do that, we chose the \ep parameter as our reference to determine the spectral change points since the COMPT model fits more than 98\% of the spectra and another parameter in the model (i.e., the power-law index) does not vary significantly within individual events. 

\begin{figure}[htbp]
    \centering
    \epsscale{1.15}
    \includegraphics[width = 14 cm, trim=70 280 40 265, clip]{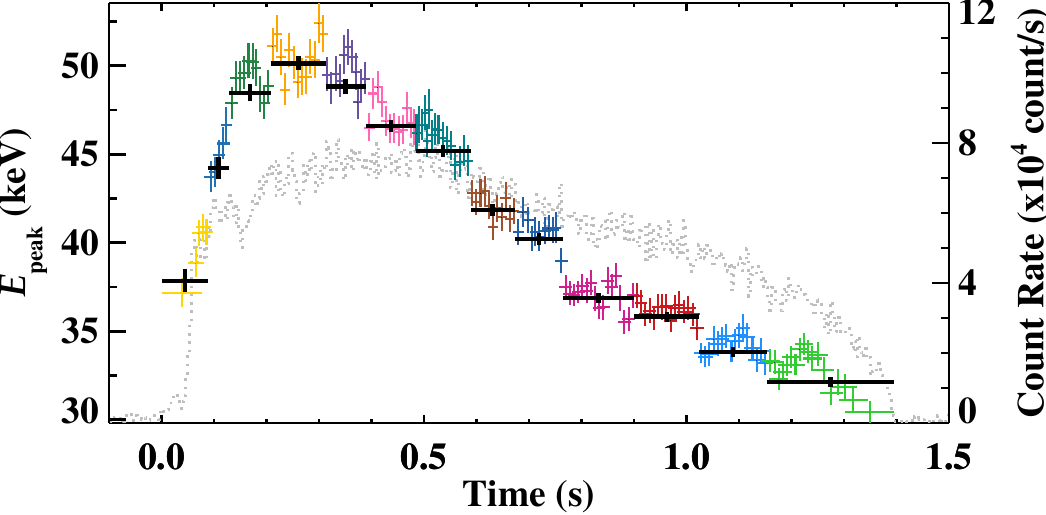}
       \caption{Light curve of an \sgr burst detected on 2021 September 10 at 00:45:46.875 UTC as seen with the brightest detector (n8) is shown with grey dotted lines (right axis). Via $k$-means clustering, 160 overlapping time segments and their corresponding $E_{\rm peak}$ values (the colored data points with asymmetric $E_{\rm peak}$ error bars) yielded 13 spectrally distinctive clusters, each of which is shown with a different color. The black data points show the $E_{\rm peak}$ values with asymmetric error bars that are obtained from the COMPT fit of the data extracted from the time spanned by these 13 clusters in the second stage of spectral analysis.}
    \label{kmeans_plot}
\end{figure}

For obtaining significant change points in the sequential \ep domain, we employed the $k$-means clustering method \citep[][]{scikit-learn} using the midpoints of the overlapping time segments and their corresponding \ep values. With the clustering, we were able to combine the consecutive time segments that yielded similar \ep values considering their errors. Thus, we increased the statistical reliability of the spectral parameters in the second stage of spectroscopy and emphasized the \ep evolution of the burst explicitly by bringing the spectral change points to the fore. In Figure \ref{kmeans_plot}, we present an example of how overlapping time segments are grouped with $k$-means clustering for the clustering-based time-resolved spectral analysis, which reveals the spectral evolution throughout the burst, independent of initial time binning. The details of how we applied the $k$-means clustering to our data can be found in Appendix \ref{appB}.


After the $k$-means clustering, the grouped time segments still overlap since they are clusters of overlapping time segments (see how the last colored time segment of a cluster overlaps with the first colored time segment of the following cluster in  Figure \ref{kmeans_plot}). Therefore, we defined the end of the cluster time interval to be the time at which half of the counts within the overlapping interval were accumulated (the black data points in Figure \ref{kmeans_plot} show the non-overlapping time intervals). After finalizing the cluster time intervals, we checked the total photon counts in each time interval before subjecting them to the second-stage analysis; We found that when there are sharp changes in the spectral parameters, a non-overlapping cluster can only contain 900 (or even less in a few cases) counts. Since we confirmed that this was still statistically sufficient for the second stage of our spectral investigations, we included all spectra of non-overlapping time segments with at least 900 counts in our study. In the few cases where the burst counts remained below this level, we combined the time segments if they are adjacent, and if they are isolated single segments, we combined them with an adjacent segment with a closer \ep value.

In the end, we obtained 287 spectrally distinctive time segments for the clustering-based time-resolved spectral investigations of 51 bursts. In Table \ref{burst_table}, we list the times of these 51 bursts, along with the detectors used in spectral analysis and event duration, as well as the number of overlapping and non-overlapping time segments.

\startlongtable
\begin{deluxetable*}{cccccc}
\tablecaption{List of 51 \sgr bursts included in our sample}
\tablehead{
\colhead{Event Time (UTC)$^a$} & \colhead{Time (\emph{Fermi} MET)} & \colhead{Detectors$^b$} & \colhead{Duration$^{c}$}  & \twocolhead{Number of Time Segments}  \\
\colhead{(YYMMDD hh:mm:ss)} & \colhead{(s)} & \colhead{}   & \colhead{(s)}  & \colhead{Overlapping}  & \colhead{Nonoverlapping} }
\tabletypesize{\footnotesize}
\startdata
\hline
160518 09:09:23.800 & 485255367.890 & 1,\textbf{9},10 & 0.161 & 13 & 3\\
160520 05:21:33.483 & 485414497.487 & 6,\textbf{7},8  & 0.108 & 9  & 3\\
160520 21:42:29.322 & 485473353.323 & 
9,\textbf{10}   & 0.268 & 7  & 3\\
160620 15:16:34.838 & 488128598.842 & 6,\textbf{7},8  & 0.224 & 21 & 6\\
160623 21:20:46.404 & 488409650.404 & 6,\textbf{7},9  & 0.488 & 32 & 8\\
160626 13:54:30.722 & 488642074.718 & \textbf{6},7,9  & 0.840 & 90 & 15\\
160721 09:36:13.665 & 490786577.665 & 6,\textbf{7},8  & 0.176 & 17 & 5\\
191104 10:44:26.230 & 594557071.231 & 0,\textbf{1},2  & 0.188 & 9  & 3\\
191105 06:11:08.595 & 594627073.579 & 3,\textbf{4},8  & 0.808 & 29 & 9\\
200410 09:43:54.273 & 608204639.277 & 
\textbf{4}      & 0.168 & 13 & 4\\
200427 18:34:05.700 & 609705250.708 & 0,1,\textbf{9}  & 0.404 & 12 & 3\\
200427 18:36:46.007 & 609705411.006 & 0,1,\textbf{9}  & 0.363 & 7  & 3\\
200427 19:43:44.537 & 609709429.537 & 3,\textbf{7},8  & 0.436 & 18 & 5\\
200427 20:15:20.582 & 609711325.581 & 1,9,\textbf{10} & 1.287 & 94 & 16\\
200427 21:59:22.527 & 609717567.528 & 
2,\textbf{10}   & 0.212 & 7  & 2\\
200428 00:24:30.311 & 609726275.311 & 4,7,\textbf{8}  & 0.236 & 13 & 5\\
200428 00:41:32.148 & 609727297.148 & 3,\textbf{6},7  & 0.436 & 5  & 2\\
200428 00:44:08.209 & 609727453.210 & 3,\textbf{6},7  & 1.276 & 28 & 7\\
200428 00:46:20.179 & 609727585.179 & \textbf{6},7,9  & 0.852 & 19 & 5\\
200429 20:47:27.860 & 609886052.860 & 4,7,\textbf{8}  & 0.420 & 17 & 5\\
200503 23:25:13.437 & 610241118.417 & \textbf{3},6,7  & 0.212 & 6  & 2\\
200510 21:51:16.278 & 610840281.278 & \textbf{10},11  & 0.424 & 15 & 4\\
210129 07:00:00.973 & 633596405.966 & 1,2,\textbf{5}  & 0.208 & 19 & 6\\
210129 10:35:39.918 & 633609344.918 &
4,\textbf{5}    & 0.308 & 21 & 6\\
210130 17:40:54.743 & 633721259.652 & 1,\textbf{2},5  & 0.256 & 10 & 4\\
210216 22:20:39.572 & 635206844.573 & 0,\textbf{1},3  & 0.344 & 12 & 4\\
210707 00:33:31.632 & 647310816.633 & 9,\textbf{10},11& 0.144 & 10 & 4\\
210710 20:26:04.407 & 647641569.407 & 9,\textbf{10},11& 0.220 & 16 & 4\\
210805 00:08:56.006 & 649814941.006 & 7,\textbf{8},11 & 0.448 & 12 & 3\\
210910 00:45:46.874 & 652927551.875 & 7,\textbf{8},11 & 1.396 & 160& 13\\
210911 05:32:38.611 & 653031163.611 & 6,\textbf{7},8  & 0.308 & 6  & 2\\
210911 13:28:54.950 & 653059739.951 & 6,\textbf{7},8  & 0.180 & 13 & 4\\
210911 15:06:43.187 & 653065608.188 & \textbf{6},7,9  & 0.404 & 38 & 11\\
210911 15:15:25.373 & 653066130.373 & 6,\textbf{9},10 & 1.176 & 91 & 16\\
210911 15:17:45.288 & 653066270.288 & 6,\textbf{9},10 & 0.944 & 31 & 7\\
210911 17:01:09.675 & 653072474.675 & 
\textbf{9},10   & 1.560 & 84 & 14\\
210911 20:22:58.772 & 653084583.772 & 
9,\textbf{10}   & 1.404 & 7  & 2\\
210912 12:19:20.431 & 653141965.431 & 
9,\textbf{10}   & 0.492 & 5  & 2\\
210912 20:16:10.382 & 653170575.381 & 
9,\textbf{10}   & 0.983 & 26 & 6\\
210912 23:19:32.042 & 653181577.043 & 
9,\textbf{10}   & 0.296 & 6  & 2\\
211008 15:57:46.393 & 655401471.394 & \textbf{6},7,9  & 0.360 & 17 & 5\\
211224 03:42:34.341 & 662010159.341 & 1,3,\textbf{5}  & 1.300 & 132& 16\\
211229 16:41:26.190 & 662488891.191 & 0,\textbf{1},3  & 0.308 & 9  & 3\\
220111 17:05:55.630 & 663613560.638 & 0,1,\textbf{3}  & 0.352 & 19 & 5\\
220112 08:39:25.279 & 663669570.275 & 0,\textbf{1},2  & 1.036 & 50 & 10\\
220112 19:58:04.026 & 663710289.027 & 0,1,\textbf{3}  & 0.744 & 12 & 3\\
220114 16:08:43.298 & 663869328.298 & 0,\textbf{1},2  & 0.568 & 11 & 3\\
220115 07:05:44.753 & 663923149.801 & 3,\textbf{4},5  & 0.688 & 9  & 2\\
220115 08:25:56.217 & 663927961.173 & 0,\textbf{3},4  & 0.684 & 5  & 2\\
220115 17:21:59.282 & 663960124.227 & 1,2,\textbf{5}  & 0.348 & 13 & 4\\
220116 14:09:38.568 & 664034983.565 & 0,\textbf{1},5  & 0.698 & 18 & 6\\
\enddata
\tablecomments{\\
$^{a}$ The bursts in 2016, in 2019$-$2020, and in 2021$-$2022 are taken from \citet{Lin2016}, \citet{Lin2020}, and Lin et al.~(in preparation), respectively.\\
$^{b}$ Unblocked NaI detectors used in spectral analysis. The brightest detectors shown in bold are used to determine the start and stop times of time segments for the extraction of spectra.\\
$^{c}$ Bayesian Block Duration. Event times (both UTC and MET) represent the Bayesian Block duration start times of the bursts.
}
\end{deluxetable*}\label{burst_table}


\subsection{Clustering-Based Time-Resolved Spectroscopy}

We performed a detailed analysis of the 287 spectra accumulated from distinct time segments with the three continuum models (COMPT, BB+BB, and MBB-RCS). Based on these results, we determined the preferred model(s) for each segment using their BIC values. We found that COMPT is a preferred model for 279 spectra (See Table \ref{photon_models}), out of which COMPT is the only preferred model for 83 spectra, COMPT \& BB+BB are equally preferred for 98 spectra, and COMPT \& MBB-RCS are equally preferred for 58 spectra. As for the thermal models, BB+BB model is statistically preferred for 145 (51\%) and MBB-RCS model for 98 (34\%) out of 286 spectra (one spectrum is excluded due to unacceptable fit statistics; see below about the goodness of fit test). Finally, all three models are equally preferred for 40 spectra (14\%). To demonstrate the spectral shapes of these three continuum models, we present an exemplary count spectrum that nearly equally fits well with all three models in Appendix \ref{appC} (Figure \ref{cnt_spec_all}).


\begin{center}
\begin{table}[ht!]
\tabletypesize{\footnotesize}
\caption{Number of non-overlapping time segments per preferred photon models}
    \begin{tabular}{|c|c|c|c|c|c|}
        \hline
        \textbf{Preferred Models$^a$}&\textbf{COMPT}&\textbf{BB+BB}&\textbf{MBB-RCS}&\textbf{All}&\textbf{Total Number ($\%$)}\\
        \hline
        COMPT (3) & 83 & 98 & 58 & \multirow{3}*{40} & 279 (97.6$\%$) \\ \cline{1-4} \cline{6-6}                   
        BB+BB (4) & 98& 7 & 0 & & 145 (50.7$\%$) \\ \cline{1-4} \cline{6-6}              
        MBB-RCS (2)& 58 & 0 & 0 & & 98 (34.3$\%$) \\
        \hline
    \end{tabular}
    \tablecomments{$^{a}$ Number of model parameters is indicated in parentheses.}
\label{photon_models}
\end{table}
\end{center}

\begin{figure}[htbp]
    \centering
    \epsscale{1.15}
    \includegraphics[width = 14 cm, trim=75 280 40 265, clip]{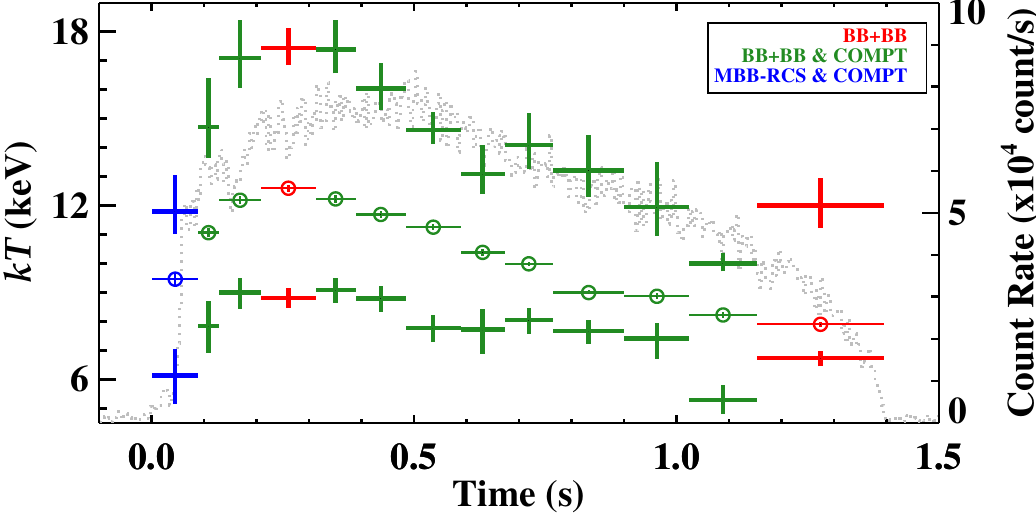}
       \caption{Blackbody temperature evolution for a burst, throughout which a thermal model is preferred. The light curve is shown with grey dotted lines (right axis; the same event as Figure\,\ref{kmeans_plot}). Thick data points show the \lkt and \hkt parameters of BB+BB while thin data points with circles show \mkt of MBB-RCS. The color code represents the preferred model(s) for each time segment based on $\Delta$BIC. The $\Delta$BIC between BB+BB and the other two models is only slightly above 10 for the first time segment, which is shown with blue color.
       }
    \label{thermal_plot}
\end{figure}

Out of the seven spectra that favor only BB+BB, we found $\Delta$BIC $\sim 30-40$ for four of them (two segments each from two bursts), meaning that the BB+BB model is definitely preferred over the other two models. Interestingly, the remaining time segments of these two bursts also favor thermal models (i.e., BB+BB and/or MBB-RCS), besides COMPT. In Figure \ref{thermal_plot}, we present the spectral evolution of thermal model parameters from one of these two bursts. We also present the time evolution of photon flux distribution for the first seven segments of this event in Appendix \ref{appC} (Figure \ref{cnt_spec_time}).


To ensure that these preferred models provide statistically acceptable fits to the spectra, we also evaluated the goodness of fits for the preferred models using C-stat, following \citet{Kaastra2017}. In doing so, we used the C-stat values excluding 30$-$40 keV to avoid the contribution from the iodine K-edge, which could affect the statistics for bright events\footnote{\url{https://fermi.gsfc.nasa.gov/ssc/data/analysis/GBM_caveats.html}}. To this end, we computed\footnote{Using the Python package \url{https://github.com/abmantz/cstat/blob/python/cashstatistic/cashstatistic.py}} the expected C-stat ($C_{\rm e}$) and its variance ($C_{\rm v}$) based on model predictions and determined whether each model fit remained within the 3$\sigma$ level\footnote{$\sigma = (\text{C-stat} - C_{\rm e}) / \sqrt{C_{\rm v}}$} of its expected C-stat distribution. We found that for only one spectrum (the second time segment of the burst at 608204639.277 MET), all three models yielded C-stat values with $>3\sigma$ deviation; this is due to the counts in $< 10$\,keV being lower than what is expected from these three models. Therefore, we excluded this time segment from our analysis and evaluated 286 out of 287 time segments from the 51 bursts in what follows below. Note that the numbers given in Table \ref{photon_models} exclude this time segment.

\begin{figure}[htbp]
    \centering
    \epsscale{1.15}
    \includegraphics[width = 14 cm, trim=70 150 100 265, clip]{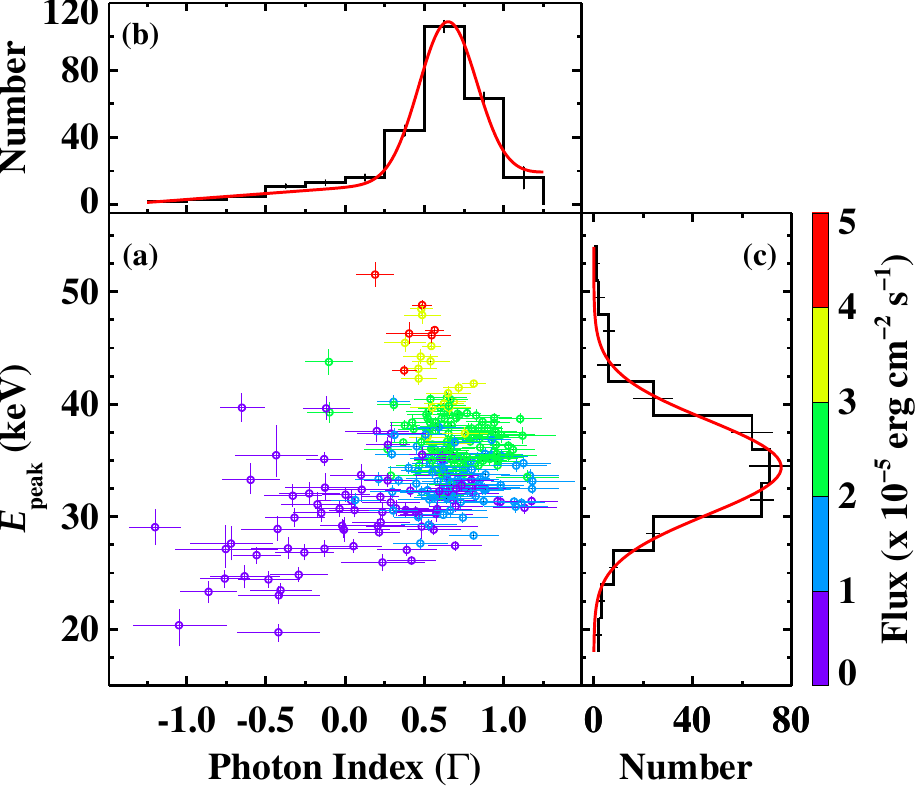}
       \caption{(a) The scatter plot of $E_{\rm peak}$ vs. $\Gamma$ parameters of 279 spectra that can be described well with COMPT. The colors indicate the energy flux values (8$-$200\,keV). (b) The distribution of $\Gamma$ values with the best fit model (Linear + Gaussian) curve is shown in red. (c) The distribution of $E_{\rm peak}$ values with the best fit Gaussian function overlaid in red.
}
    \label{nonthermal_fits}
\end{figure}

In Figure \ref{nonthermal_fits} panel (a), we present the scatter plot of the 279 pairs of photon index ($\Gamma$) and $E_{\rm peak}$ values of the COMPT model. Note that the plot is color-coded based on the energy flux in 8$-$200\,keV. We find that the distribution of  $\Gamma$ is asymmetric, described best by a Gaussian with an underlying first-order polynomial (see panel (b) of Figure \ref{nonthermal_fits}). $\Gamma$ values above 0.25 follow a Gaussian with a mean value of 0.65$\pm$0.02 and $\sigma$ = 0.18$\pm$0.02. However, $\Gamma$ values below 0.25 form an excess above the Gaussian tail, which can be described with a first-order polynomial of (7.08$\pm$1.95)$\Gamma$ + (10.13$\pm$1.76). It is clear from Figure \ref{nonthermal_fits} that those $\Gamma$ values forming the excess are obtained from time segments with the lowest flux in our sample (below $1 \times 10^{-5}$ erg cm$^{-2}$ s$^{-1}$). These time segments with low flux values generally correspond to the first or last time segments of the bursts as expected. On the other hand, time segments with higher flux values yield $\Gamma$ values mostly larger than 0.25, and those are the values whose distribution is consistent with a Gaussian. It is also important to note that the time segments with the highest flux values (above $4~\times$ 10$^{-5}$ erg cm$^{-2}$ s$^{-1}$) yield $\Gamma$ values in a very narrow range from 0.2 to 0.55. 


As for the corresponding $E_{\rm peak}$ values, they range from $\sim$20 to $\sim$52\,keV, and their distribution follows a Gaussian shape with a mean of 34.39$\pm$0.26\,keV and a width, $\sigma$ = 4.21$\pm$0.20 (see panel (c) of Figure \ref{nonthermal_fits}). Moreover, there appears a positive correlation between $E_{\rm peak}$ and flux. To quantify the correlation, we computed Spearman's rank order correlation coefficients ($\rho$ and chance probability, $P$) for $E_{\rm peak}$ and flux using a bootstrap method; we generated 10000 data sets by taking into account the uncertainties of $E_{\rm peak}$ and flux, and calculated the correlation coefficients for each data set. From the distribution of these coefficients, we obtained the mean and 1$\sigma$ confidence interval of $\rho$ and $P$. We found $\rho=0.81 \pm 0.01$ and $P < 10^{-63}$ for $E_{\rm peak}$ and flux, which lends support to the observed positive correlation.

\begin{figure}[htbp]
    \centering
    \epsscale{1.15}
    \includegraphics[width = 14 cm,  trim=70 142 100 265, clip]{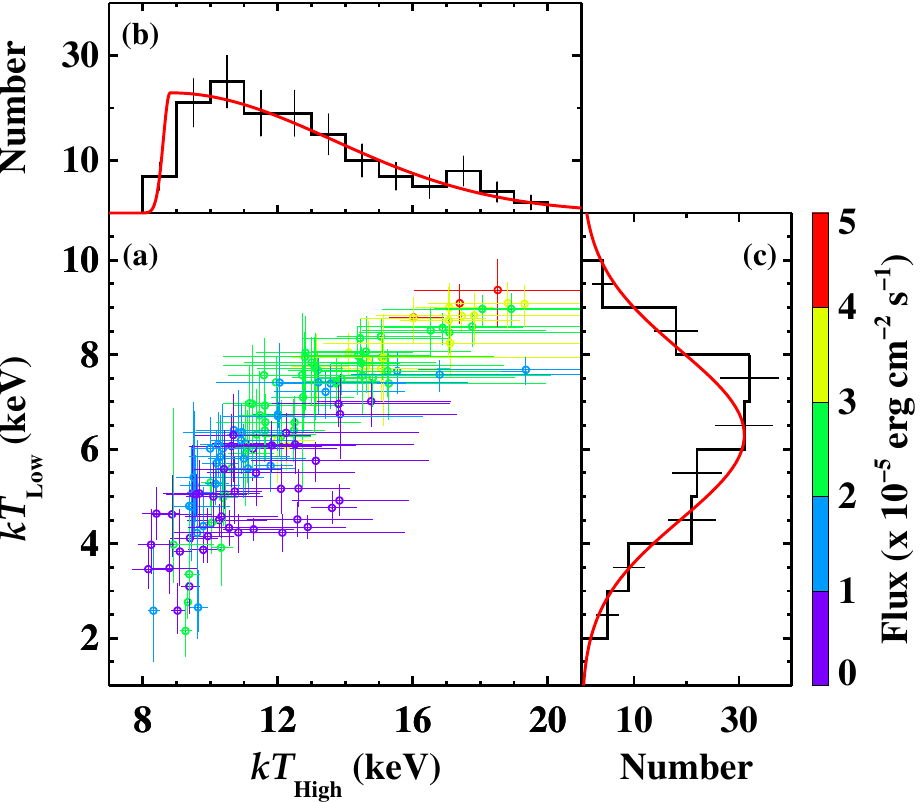}
       \caption{(a) The scatter plot of \lkt vs. \hkt parameters of 145 spectra that can be described well with BB+BB. The colors indicate the energy flux values. (b) The distribution of \hkt with the best fit model (two-sided Gaussian) curve is shown in red. (c) The distribution of \lkt values with the best fit Gaussian function overlaid in red.
}
    \label{thermal_fits}
\end{figure}

For the 145 BB+BB-preferred spectra, we present the scatter plot of low BB (\lktnosp) and high BB (\hktnosp) temperatures in Figure \ref{thermal_fits}. It is again color-coded based on the energy flux in the 8$-$200\,keV. We find a positive correlation between the two parameters ($\rho=0.67 \pm 0.04$ and $P < 10^{-17}$). Overall, lower temperatures are associated with low flux levels (purple and blue data points in panel (a) of Figure \ref{thermal_fits}), while higher temperatures correspond to high flux values (red and yellow data points). The intermediate flux values, however, have a much wider BB+BB temperature range. In terms of parameter distributions, we find that a best Gaussian fit to the distribution of \lkt yields a mean of 6.30$\pm$0.17\,keV with a sigma of 1.80$\pm$0.15 (see panel (c) of Figure \ref{thermal_fits}). On the other hand, our criterion that \hkt parameter must be larger than \lkt parameter forces the distribution of \hkt to be asymmetrical (see panel (b) of Figure \ref{thermal_fits}). Therefore, the best fit to the distribution of \hkt results in a two-sided Gaussian fit. The distribution has a mean of 8.81$\pm$2.74\,keV with $\sigma_{\rm left}$ = 0.2$\pm$1.8 and $\sigma_{\rm right}$ = 4.83$\pm$1.4. 


\begin{figure}[htbp]
    \centering
    \epsscale{1.15}
    \includegraphics[width = 8.5 cm, trim=90 149 225 340, clip]{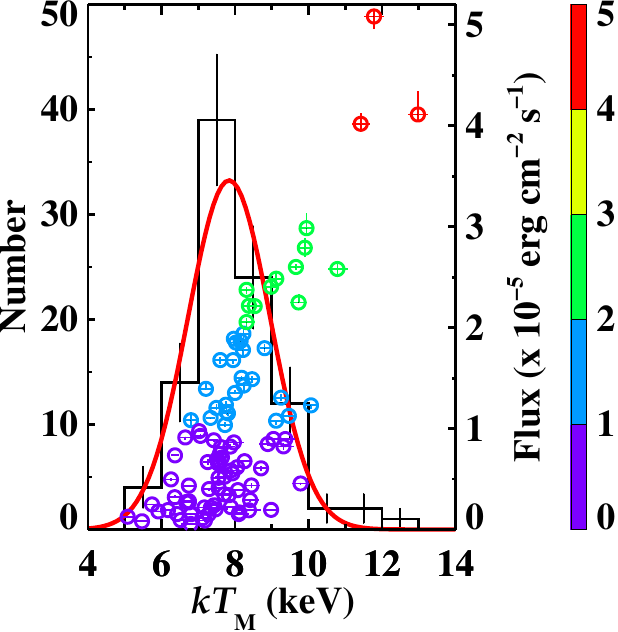}
       \caption{The distribution of \mkt parameter of 98 spectra that favor MBB-RCS, with the best-fit Gaussian function overlaid in red (left axis). The scatter plot of energy flux vs. \mkt is also shown (right axis). The colors indicate the energy flux values.
}
    \label{lbb_fit}
\end{figure}

Finally, for the 98 MBB-RCS-preferred spectra, the distribution of temperature (\mktnosp) lies between 5$-$13\,keV, which remains in between \lkt and \hkt distributions of the BB+BB model. In Figure \ref{lbb_fit}, we present the \mkt parameter distribution of the MBB-RCS model. The distribution is described well with the normal distribution, which peaks at 7.84$\pm$0.12\,keV with $\sigma$ = 1.12$\pm$0.1. As in the BB+BB model, we also observe a positive correlation between energy flux and temperature in this model ($\rho=0.62 \pm 0.02$ and $P < 10^{-11}$); on average, higher \mkt values correspond to higher flux values.


\section{Discussion}\label{sec:resndis}

In this study, we applied a novel approach for the first time in magnetar bursts research: a clustering-based time-resolved spectroscopy.  We performed high time-resolved spectral analysis of the brightest \sgr bursts by first dividing them into the sequentially-overlapping shortest time intervals to clearly reveal the spectral evolution during the short magnetar bursts. We then identified the significant spectral change points throughout the bursts using the spectral analysis results of these overlapping time segments through a machine learning based clustering approach and completed the second round of spectroscopy with the data extracted from the time intervals between these change points. In the end, we obtained 287 non-overlapping sequential time segments from 51 bursts. Out of 287 spectrally-distinguishing time intervals, 207 arise from the brightest half of our burst sample, indicating the existence of a quite significant spectral evolution in the brightest magnetar bursts.

\citet{Lin2016,Lin2020} studied time-integrated spectral properties of 275 \sgr bursts between 2014 and 2020 observed with \ferminosp. They found that $\sim$62\% and $\sim$38\%  could be described well with the BB+BB and COMPT models, respectively. On the contrary, our clustering-based time-resolved spectral analysis revealed that only $\sim$51$\%$ can be fit with BB+BB while nearly all time segments ($\sim$98$\%$) of the brightest bursts could be represented with COMPT. Note the fact that our burst sample selected from these 275 bursts are the brightest ones, based on our criterion as explained in Section \ref{sec:info}; therefore, it would be a fairer comparison with our results if we exclude dim bursts from their statistics. In that case, the accepted fit percentages for their samples (178 bursts) increase to $\sim$96$\%$ and $\sim$58$\%$ with the BB+BB and COMPT models, respectively, which still are quite different from our results found here. Besides the difference between time-integrated and time-resolved analysis, such a difference in model preferences of the two studies might arise from the fact that our sample includes also \sgr bursts from 2021-2022 active episode. However, our statistics remained nearly the same when we excluded these bursts and evaluated only the bursts before 2021, and even when we look at the percentages of statistically acceptable fits (instead of BIC-based preferences), COMPT is a better-describing model than BB+BB. We thus conclude that the time-resolved spectral results are different from those of time-integrated spectral investigations. This is not surprising since we observe significant spectral variations throughout bursts and superposition of spectra with varying \textit{E}$_{\rm peak}$ of COMPT could mimic the spectral energy distribution of BB+BB or results in the distribution that deviates from a single-component COMPT model.  Besides, the time-integrated spectra naturally have more counts even at higher energies, with which the two-component model parameters are better determined.

\subsection{Spectral Parameters-Flux-Area Correlations}\label{sec:sub_sec1}

As for the model parameters of the time-integrated spectroscopy of \sgr bursts between 2014$-$2016, those that fit well with the COMPT model have $E_{\rm peak}$ values range between $\sim$25 and 40\,keV with a Gaussian mean of 30.4$\pm$0.2\,keV \citep{Lin2016}.  The later 2019$-$2020 bursts have a wider range of $E_{\rm peak}$  ($\sim$10$-$40\,keV) with a slightly lower mean of 26.4$\pm$0.6\,keV \citep{Lin2020}. In comparison, our clustering-based time-resolved spectroscopy yields a slightly higher energy range of $\sim$20$-$52\,keV and a higher mean $E_{\rm peak}$ of 34.4$\pm$0.3\,keV. We again note here that our burst sample includes bursts detected in the 2021 and 2022 active episodes and met our brightness criterion; when we compare only the bright events between 2016 and 2020, the time-resolved $E_{\rm peak}$ in our sample is still harder.

\begin{figure}[htbp]
    \centering
    \epsscale{1.15}
    \includegraphics[width = 15 cm,  trim=13 347 12 110, clip]{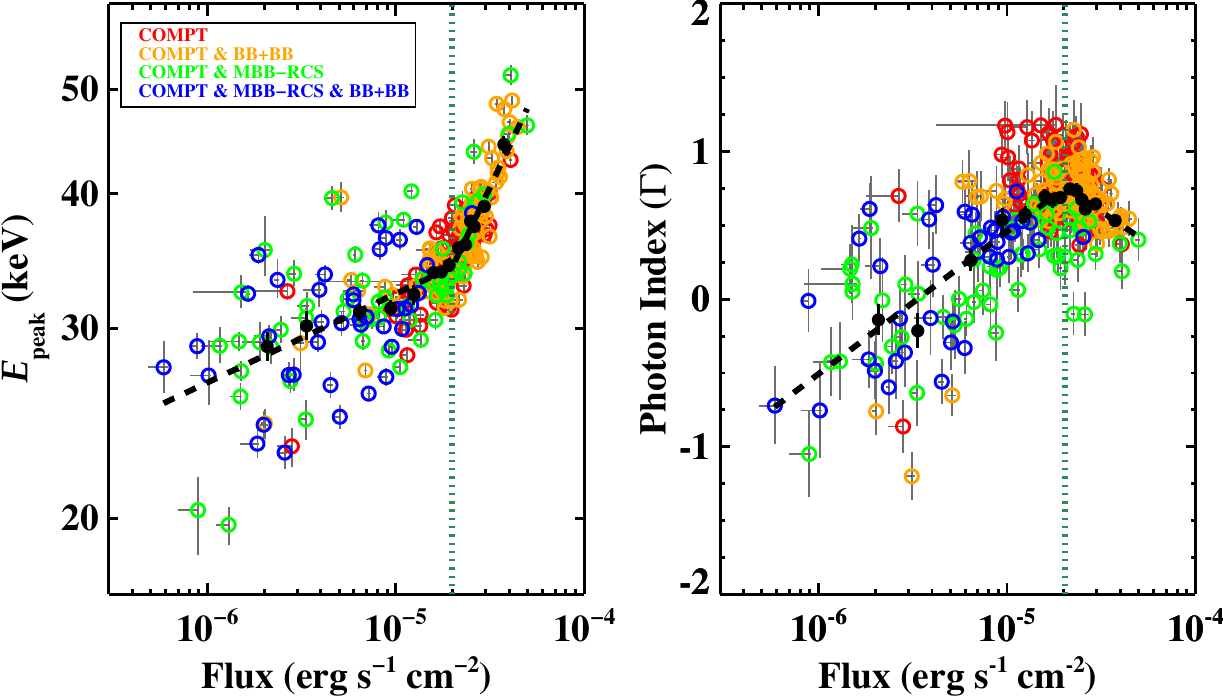}
       \caption{The scatter plot of $E_{\rm peak}$ vs. flux \textbf{(left panel)} and photon index vs. flux \textbf{(right panel)} of the COMPT model fits. The color code shows the preferred photon model(s) based on BIC values. The black dots represent the weighted means of consecutive groups, each with 20 data points. The black dashed lines show the BPL fits to the relation between the weighted means of $E_{\rm peak}$ and flux, and between the weighted means of photon index and flux, respectively. The vertical dotted lines in both panels show the flux breaks, which are consistent with each other within their errors.
}
    \label{epeak_index_flux}
\end{figure}

Previously, time-integrated spectral analysis of \sgr bursts and other prolific magnetar bursts (SGR\,J1550$-$5418) showed that $E_{\rm peak}$ is correlated with the energy flux or fluence described by a power law or a broken power law \citep{von12, Lin2016, Lin2020}. We present in Figure \ref{epeak_index_flux} (left panel), the $E_{\rm peak}$ vs. flux plot for our time-resolved results; the correlation was revealed more clearly as a result of our time-resolved spectroscopy (Spearman's rank correlation coefficient, $\rho=0.81 \pm 0.01$, $P < 10^{-63}$). To better quantify the relation between $E_{\rm peak}$ and flux, we fit the trend as follows: Since flux errors are small ($\Delta F/F\lesssim0.015$), we grouped the data in the flux domain such that each group would include 20 data points. For each group, we computed the weighted mean flux and $E_{\rm peak}$, as well as 1$\sigma$ uncertainty of $E_{\rm peak}$. Modeling the grouped trend with a single power law model (PL) yields an unacceptable fit ($\chi^2$/dof = 117.4/12). A fit with a broken power law model (BPL), on the other hand, results in statistically acceptable representation ($\chi^2$/dof = 15.7/10), yielding a break at the flux of (2.00$\pm$0.05) $\times 10^{-5}$ erg cm$^{-2}$ s$^{-1}$, and positive indices of 0.08$\pm$0.01 and 0.37$\pm$0.03 before and after the break, respectively (see Figure \ref{epeak_index_flux} left panel).

On the other hand, we found that the range of the COMPT photon index ($\Gamma$) parameter of our time-resolved investigation is consistent with the time-integrated spectroscopy although their distributions are quite different. The $\Gamma$ distribution of time-integrated spectral analysis runs from $-$1.5 to 1 and follows a Gaussian with a mean of $\sim$0 \citep{Lin2016, Lin2020}. In the case of our time-resolved spectroscopy, the distribution has a tail, best described with a Gaussian with an underlying first-order polynomial. The positive $\Gamma$ values above 0.25 are distributed like a Gaussian with a mean at 0.65$\pm$0.02, while $\Gamma$ values below 0.25 form an excess above the Gaussian tail.  We found that these indices ($\Gamma < 0.25$) were obtained from the spectra with the lowest flux values in our sample ($< 1 \times 10^{-5}$ erg cm$^{-2}$ s$^{-1}$). Moreover, the spectra with the highest flux values in our sample ($> 4 \times 10^{-5}$ erg cm$^{-2}$ s$^{-1}$) yield $\Gamma$ values in a very narrow range with a weighted mean of 0.47$\pm$0.03. There exists a positive correlation between $\Gamma$ and flux up to a certain flux level that coincides with the flux break of $E_{\rm peak}$ vs. flux (see the right panel of Figure \ref{epeak_index_flux}). The index and flux are anti-correlated after that point. The change in trend between index and flux is also indicated in Figure \ref{nonthermal_fits}: the photon index increases with increasing flux up to about $2 \times 10^{-5}$ erg cm$^{-2}$ s$^{-1}$ (blue and purple data points in the panel (a) of Figure \ref{nonthermal_fits}). Then, the index starts to decrease with increasing flux. In modeling this trend, we followed a similar approach as in the case of $E_{\rm peak}$ vs. flux: We obtained weighted mean $\Gamma$ values and their 1$\sigma$ uncertainties for the groups of 20 flux values and used a broken log-linear function to fit, yielding $\chi^2$/dof = 10.1/10. We find the break at the flux of (2.04$\pm$0.09)$\times 10^{-5}$ erg cm$^{-2}$ s$^{-1}$, that is consistent with the $E_{\rm peak}$ vs. flux case. The slope of 0.97$\pm$0.09 before the break changes to $-$0.86$\pm$0.17 afterwards (see the right panel of Figure~\ref{epeak_index_flux}).

\citet{Younes14} performed time-resolved spectral investigations of bursts from another prolific magnetar, SGR\,J1550$-$5418, also observed with \ferminosp. They also found that the $E_{\rm peak}$ vs.~flux relation is described better with a BPL rather than a single PL. This break point is at the flux of $\sim$ 1 $\times$ 10$^{-5}$ erg cm$^{-2}$ s$^{-1}$, which is about a half the break value in flux that we found for \sgrnosp. Unlike our findings, the $E_{\rm peak}$ vs. flux correlation of SGR\,J1550$-$5418 is negative at low flux values, while it is positive after the flux break. Moreover, $\Gamma$ remains constant at $\sim$ $-$0.8 up to the flux break, then follows the same positive trend as seen between $E_{\rm peak}$ and flux. Similar dual relation between the $E_{\rm peak}$ and flux with a break at the flux of $\sim$ 1 $\times$ 10$^{-5}$ erg cm$^{-2}$ s$^{-1}$ was also reported in the time-resolved spectroscopy of five bright bursts from yet another magnetar, SGR\,J0501+4516 \citep{Lin2011}.

For the BB+BB model, earlier time-integrated spectral studies of \sgr bursts yielded parameters of $\sim$2$-$8\,keV for \lkt with a mean of 4.5\,keV, and $\sim$8$-$20\,keV for \hkt with a mean of 11\,keV \citep{Lin2016,Lin2020}. Our clustering-based time-resolved spectroscopy using the BB+BB model results in similar parameter range and distribution for \hktnosp. However, our \lkt parameter reaches up to 10\,keV with a larger mean of 6.3$\pm$0.2\,keV. Moreover, unlike the time-integrated spectral analysis, our study reveals a positive correlation between \lkt and \hkt ($\rho=0.67 \pm 0.04$, $P < 10^{-17}$; see panel (a) of Figure \ref{thermal_fits}). 

\begin{figure}[htbp]
    \centering
    \epsscale{1.15}
    \includegraphics[width = 15 cm,  trim=10 352 10 100, clip]{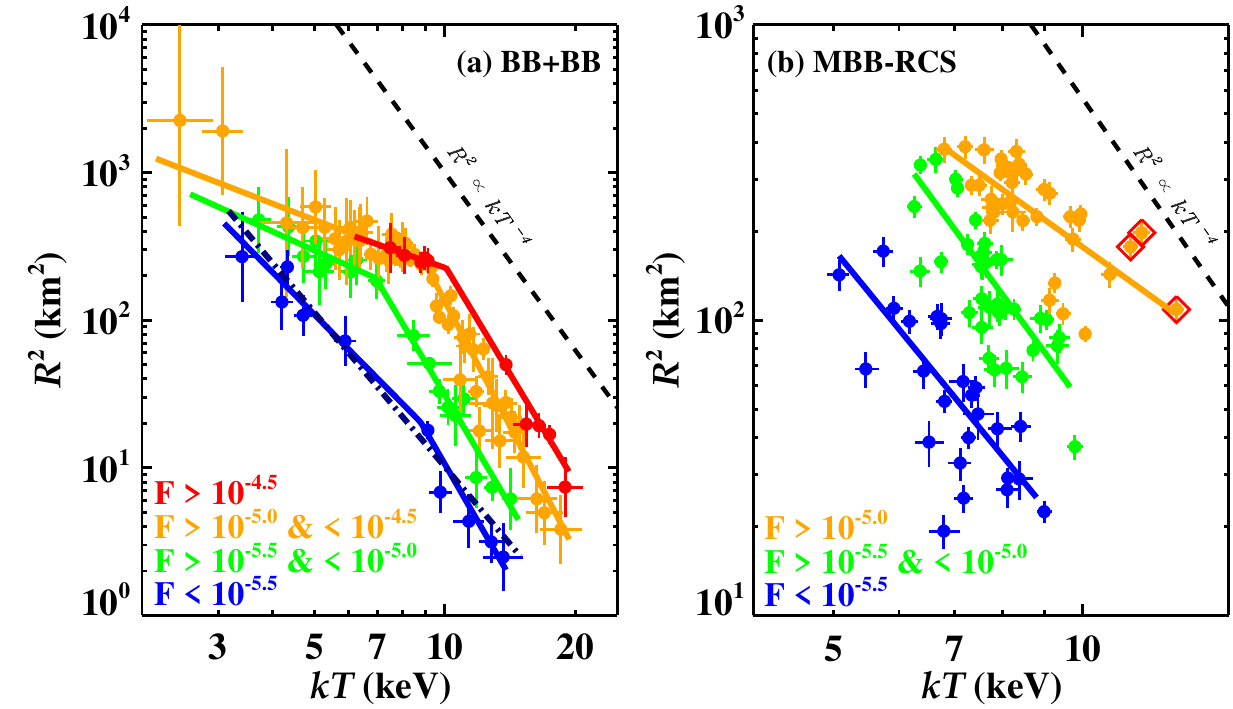}
       \caption{\textbf{[Left]} Flux color-coded plot of $R^{2}$ vs. $kT$ for 145 BB+BB spectra. Each data point represents the weighted means of $R^{2}$ and $kT$ of three time segments only for display purposes. Solid lines show BPL fits. The dark blue dash-dotted line indicates PL fit to the lowest flux group. The lowest flux group is fitted almost equally well with both a single PL and BPL. The black dashed line indicates $R^2 \propto kT^{-4}$. \textbf{[Right]} Flux color coded scatter plot of $R^{2}$ vs. $kT$ for 98 spectra favoring MBB-RCS. Solid lines represent PL fits. The dashed line indicates $R^2 \propto kT^{-4}$. Note that we combined the data points of the two highest flux groups for MBB-RCS since only three spectra in the highest flux group ($F > 10^{-4.5}$ erg cm$^{-2}$ s$^{-1}$) favor the MBB-RCS model, which are shown with red diamonds.
}
    \label{r2_kT}
\end{figure}

Based on the BB+BB parameters, we also calculated the size of blackbody emitting areas as $R^{2} = (F d^2)/(\sigma T^4)$ in km$^2$ where $F$ is the flux, $d$ is the distance to the source (here we use a distance of 9 kpc, consistent with \citealt{Lin2016, Lin2020}), $\sigma$ is the Stefan-Boltzmann constant, and $T$ is the temperature in Kelvin. Our study revealed that the relation between $T$ and $R^{2}$ varies with the burst flux, similar to the findings of time-resolved spectral analysis of SGR\,J1550$-$5418 bursts \citep{Younes14}. Therefore, following \citet{Younes14}, we divided our 145 spectra that favors BB+BB model into four groups based on energy flux (in erg cm$^{-2}$ s$^{-1}$): $F < 10^{-5.5}$, $10^{-5.5} < F < 10^{-5.0}$, $ 10^{-5.0} < F < 10^{-4.5}$, and $F > 10^{-4.5}$. In the left panel of Figure \ref{r2_kT}, we present the plot of $R^2$ vs. $kT$ for the above-mentioned flux ranges. We found that the best fit for the $R^2$ vs. $kT$ relation of the three highest-flux data groups is the BPL model. However, for the lowest-flux group, both PL and BPL provide statistically acceptable fits.  Despite the difference, we observe that the overall trends in all four groups are the same: The negative correlation (slope) between $R^{2}$ and $kT$ becomes steeper after the break of the BPLs. We summarize our flux-dependent $R^{2}$ vs. $kT$ fit results in Table \ref{area_kT}. Note that these fit results are obtained with the data from all of the spectra that favor BB+BB, not with the weighted means of the data (that we show in Figure \ref{r2_kT} only for display purposes).


\begin{table}[htbp]
\begin{center}
\tabletypesize{\footnotesize}
\caption{$R^{2}$ vs. $kT$ fit parameters (PL index $\alpha$ and break energy) for various flux ranges of BB+BB and MBB-RCS models as shown in Figure \ref{r2_kT}
}
\label{area_kT}
    \begin{tabular}{c|ccc|c}

        &   &\textbf{BB+BB} &   & \textbf{MBB-RCS} \\
        
        \hline
        \hline
        
        Flux Range & $\alpha$-\lkt & $\alpha$-\hkt & $kT_{\rm break}$& $\alpha$-\mkt  \\
        (erg cm$^{-2}$ s$^{-1}$)& & &(keV) & \\  
        
        \hline
        
        $F > 10^{-4.5}$              & $-1.02 \pm 1.64$ & $-4.88 \pm 0.79$ & $10.08 \pm 1.12$ & $-2.01 \pm 0.10 ^{a}$ \\
        
        \hline
        
        $ 10^{-5.0} < F < 10^{-4.5}$ & $-1.14 \pm 0.35$ & $-5.62 \pm 0.24$ & $8.98 \pm 1.02$ 
        & $-2.01 \pm 0.10 ^{a}$ \\
        
        \hline
        
        $ 10^{-5.5} < F < 10^{-5.0}$ & $-1.36 \pm 0.81$ & $-4.99 \pm 0.47$ & $6.96 \pm 1.09$ 
        & $-3.83 \pm 0.15$ \\
        
        \hline
        
        $ F < 10^{-5.5}$             & $-3.01 \pm 0.64$ & $-5.05 \pm 0.88$ & $8.79 \pm 1.24$ & $-3.43 \pm 0.16$ \\
        & $-3.47 \pm 0.23 ^{b}$  & $--$ & $--$ & \\
        
        \hline
    
    \end{tabular}
    \end{center}
    \tablecomments{ \\
    $^{a}$ Slopes of the data points from the highest two flux groups for the MBB-RCS model are the same since they are combined due to the deficiency of time segments (only three) that favor MBB-RCS in the highest flux regime. \\ $^{b}$ A single PL fit to the data.}

\end{table}


We observe that the $kT$ range of \sgr and SGR\,J1550$-$5418 bursts are similar while $R^{2}$ for \sgr bursts do not go down below 1 km$^2$ unlike SGR\,J1550$-$5418 bursts. Still, the relation between $R^{2}$ and $kT$ has the same trend for both sources: $R^{2}$ decreasing more slowly for \lkt compared to \hkt across all flux ranges. We found that the slope between $R^{2}$ \& \lkt ($\alpha$-\lktnosp) and the slope between $R^{2}$ \& \hkt ($\alpha$-\hktnosp) for all flux groups in both studies are consistent with one another within at most 3$\sigma$ errors. Also, $kT_{\rm break}$ values are consistent with one another within errors ($\sim$7$-$9\,keV for SGR J1550$-$5418; \citealt{Younes14}). Moreover, we observed a similar relation between $R^{2}$ and \lkt: $\alpha$-\lkt becomes steeper as flux decreases. However, our investigations do not yield a decreasing trend in the $\alpha$-\hkt with decreasing flux. Furthermore, similar to SGR J1550$-$5418 bursts, a single PL model yields statistically acceptable $R^{2}$ \& $kT$ fit in the lowest flux group.

Finally, for the other thermal model, namely, the modified blackbody whose emission undergoes resonant cyclotron scattering (MBB-RCS), we observe a correlation between temperature and flux (see Figure \ref{lbb_fit}, $\rho=0.62 \pm 0.02$ and $P < 10^{-11}$). Also for the 98 MBB-RCS-preferred spectra, we calculated the corresponding emitting region ($R^2$) based on their temperature ($kT_{\rm M}$).  In the right panel of Figure \ref{r2_kT}, we present the scatter plot of $R^{2}$ vs. \mkt with color-coding based on the same flux ranges used for BB+BB $-$ except that we combined the highest two flux groups for the MBB-RCS model due to the deficiency of time segments (only three) in the highest flux regime well described by MBB-RCS.  We find that for $R^{2}$ vs. \mkt the best fit across all flux ranges is a PL, which is presented also in Table \ref{area_kT}. We also find that PL indices ($\alpha$-\mktnosp) for the lowest two flux regimes are consistent with $-4$ (i.e., as expected from $F \propto \sigma T^4$).  More importantly, $\alpha$-\mkt obtained from the highest flux spectra (\textit{F} $>$ 10$^{-5}$ erg cm$^{-2}$ s$^{-1}$, corresponding to an isotropic luminosity of 10$^{41}$ erg s$^{-1}$) is significantly different, $-$2.01 $\pm$ 0.10, than the lower-flux spectra.

\subsection{Interpretative Elements}\label{sec:sub_sec2}

We now elaborate on the interpretation of some of our results. Large Thomson opacities are expected for magnetars in outburst, quickly discernible using non-magnetic estimates that were identified by \cite{BH-2007-ApandSS}.  Setting \teq{{\cal E}_e\gtrsim L_{\gamma}/(4\pi R^2c)} as the representative kinetic energy density in radiating electrons, then if they possess a typical Lorentz factor \teq{\langle\gamma_e\rangle\sim 1}, one quickly arrives at an electron number density \teq{n_e \sim  {\cal E}_e/(m_ec^2)}, so that the non-magnetic Thomson optical depth is \teq{\taut =n_e\sigt R\gtrsim L_{\gamma}\, \sigt/(4\pi Rm_ec^3)}; this is the familiar compactness parameter.  For \teq{R\sim 10^7}cm, this yields \teq{\taut\sim 10^4} (i.e., \teq{n_e\gtrsim 10^{21}}cm$^{-3}$) for SGR bursts of typical isotropic luminosities \teq{L_{\gamma}\sim 10^{40}}erg\,s$^{-1}$, indicating optically thick, super-Eddington conditions \citep[e.g.][]{TD-1996-ApJ} that drive plasma flow along the field lines.

In Figure \ref{r2_kT} (left), we observe a significant deviation from the Stefan-Boltzmann (S-B) law for an isotropic radiation field ($R^2 \propto kT^{-4}$): We find $R^2 \propto kT^{-\alpha}$, where $\alpha\sim$1$-$1.4 for \teq{kT\lesssim 7-10}keV above the flux level of $\sim 10^{-5.5}$ erg cm$^{-2}$ s$^{-1}$.  Note that this flux corresponds to an isotropic luminosity of $3 \times 10^{40}$ erg s$^{-1}$ at a source distance of 9 kpc. The resulting broken power-law $R^2-kT$ correlations can be interpreted as a signature of the spatial extension of the active emission region, which can be presumed to be a broad, flaring flux tube.  If this tube has a transverse dimension of $R_t\sim 2-10\,$km for its cross section at the highest altitudes, one can infer a tube length of $R_l\sim 4 R^2/R_t \sim 100-500\,$km for the largest $R^2$ values.  The highly optically thick gas in the emission region will naturally cool adiabatically when moving between smaller, hotter regions near the flux tube footpoints at the stellar surface, and the high altitude, large area regions near the equatorial tube apex. This yields the spectral extension we see. If the radiating gas is locally quasi-thermal, then near the footpoints the magnetic field is high, and photospheric/atmospheric simulations \citep[see Fig.~2 of][]{Hu-2022-ApJ} of radiative transfer in this sub-cyclotronic frequency domain ($\omega \ll \omega_B = eB/\hbar c$) indicate that the radiation is quasi-isotropic.  In contrast, since the magnetic field is much lower (likely 3-4 orders of magnitude) near the tube apex, the Compton scattering radiative transfer in the local region samples the cyclotronic domain where $\omega \sim \omega_B$. This domain evinces significantly anisotropic emergent radiation fields \citep[see the right column of Fig.~2 of][]{Hu-2022-ApJ}, with a decrement of intensity over a wide range of directions that are oriented closer to the outer surface of the magnetic flux tube/photosphere; this tube surface is aligned with the local field direction that guides plasma motion and associated adiabatic cooling. Such a decrement alters the S-B law from its isotropic `$R^2 (kT)^4=$ constant' form, and lowers the perceived area in lower \textit{kT} regions at higher altitudes\footnote{As a side effect, depending on the magnetar's spin phase, the star itself may obscure a segment of the plasma tube (particularly, the footpoint where radiation becomes isotropic) even during the burst. This could introduce additional anisotropy to the emerging radiation and perceived surface area.}. The reduced values of $R^2 (kT)^4$ naturally generate the breaks apparent in Figure \ref{r2_kT}. In particular, the actual flux tube areas would rise above those the S-B estimate obtains.  Concomitantly, this decrement yields the break in the flux-$E_{\rm peak}$ correlation depicted in Figure \ref{epeak_index_flux} (left).

The extended spatial emission zone scenario suggests that a multi-blackbody fit would be preferable to just a BB+BB one. This may be the reason why the MBB-RCS model provides a good spectral fit for a significant portion of the sample studied here, in which for the first time the MBB-RCS model has been applied to a substantial sample size.  Clearly, the scattering component of the MBB-RCS picture applies outside the highly optically thick primary burst emission zone.  If the plasma density becomes high enough, as may be more likely for the highly energetic events ($L \gtrsim$ 10$^{41}$ erg s$^{-1}$), multiple scatterings of photons by the magnetospheric charges would arise, and this likely would harden the emerging spectrum. The required plasma densities would be considerably higher than those needed in resonant inverse Compton emission models \citep{BH-2007-ApandSS,Fernandez-2007-ApJ,Wadiasingh-2018-ApJ} of the persistent hard X-ray tail emission of luminosities \teq{L\sim 10^{35}}erg s$^{-1}$ \citep{Kuiper-2004-ApJ,Goetz-2006-AandA,denHartog-2008-AandA} from various magnetars. In such domains, the MBB-RCS model would need to be expanded \citep{Yamasaki20}. 

Accordingly, strong motivations exist for future detailed accounting of the altitudinal dependence of the anisotropy of the flux tube radiation field, in combination with RCS modeling in neighboring magnetospheric regions.  The RCS process would have to address both single and multiple scattering domains.  This level of sophistication is desirable for more precisely interpreting area-color correlations at different luminosity levels.  In particular, if the energies $kT\sim 7-10\,$keV of the breaks are interpreted as a loose measure of the $\hbar \omega_B$ value (when the anisotropy becomes substantial) somewhat near the tube apex, it may prove possible to constrain the active flux tube dimensions and magnetospheric locale using the area-color correlations.  Thus, our results not only encourage further employment of the MBB-RCS physical model in diverse burst samples from various sources but also highlight the need for more comprehensive modeling approaches in understanding the behavior of highly energetic magnetar flares.

\section*{acknowledgments}
We thank the referee for the thorough review and constructive comments that improved the quality of our manuscript. \"O.K., E.G, Y.K. and M.D. acknowledge the support from the Scientific and Technological Research Council of Turkey (T\"UB\'ITAK grant no. 121F266). S.Y. acknowledges the support from the National Science and Technology Council of Taiwan through grants 110-2112-M-005-013-MY3, 110-2112-M-007-034-, and 112-2123-M-001-004-.  M.G.B. thanks NASA for generous support under awards 80NSSC22K0777 and 80NSSC22K1576.


\appendix

\section{Spectral Data Extraction} \label{appA}

For each event, we calculated the burst duration (i.e., the duration over which burst spectral extraction is to be performed) using the data collected with the brightest NaI detector, which is the detector with the smallest zenith angle to the sky position of the source. In particular, we first constructed a light curve with 4 ms time resolution for the time interval of $-$10 s to 10 s with respect to the burst start time published in \citet{Lin2016, Lin2020}, using time-tagged photon data in the 8$-$200\,keV energy band. We then generated the Bayesian Block representation of the light curve. Blocks longer than 4 s were considered background blocks, and the background level was calculated by averaging the count rates of these background blocks. The blocks shorter than 4 s and higher in rates than the background level are taken as burst blocks, and the duration is calculated as the time interval from the start of the first burst block to the end of the last one \citep{Lin2013}.

For defining sequentially overlapping time segments of each burst, we first determined the background level as the average of the pre-burst time interval between $-$50 s and $-$1 s by taking the start time of Bayesian Block duration as the reference point. Then, we obtained a background-subtracted light curve with 4 ms time resolution in the energy range of 8$-$200\,keV and determined the beginning and end points of time segments for the spectral data extraction, each of which contains at least 1200 background-subtracted counts and overlaps 80\% in time with the previous one. As stated in Section \ref{sec:info}, in the case of accumulation of time segments at the peak of a burst, we reduced the overlap by 5\% incrementally until the endpoint of the subsequent time segment ends later than the end of the previous one.

We aim to investigate each burst by dividing it into as many short time intervals as possible. At the same time, we also aim for these intervals to have sufficient statistics (i.e., burst counts) for reliable spectral analysis. Therefore, setting a threshold background-subtracted counts to be included in each time segment is required. To this end, we selected a small set of bursts and extracted spectra of the time segments for each of these bursts, again with 80\% overlap, but with a set of different numbers of background-subtracted counts, namely 600, 800, 1000, 1200, 1500, 1800, and 2000. We then fit these spectra with the three models that we used in this study and checked whether the resulting model parameters are constrained within at least 2$\sigma$ level. Based on this analysis, we concluded that a threshold of 1200 background-subtracted counts was needed to ensure reliable spectral analysis results with constrained model parameters for each time segment. With this requirement of 1200 burst counts in each time segment, the minimum number of counts throughout a burst should be larger than about 2400 in order for a burst to consist of at least two non-overlapping time segments, therefore, spectral evolution could be defined.

We also checked the potential count saturation in the data, from which very bright bursts may suffer. The GBM TTE data are affected by 2.6 $\mu$s deadtime as a result of the fixed data-packet processing speed of 375 kHz on the spacecraft \citep{Meegan2009}. Therefore, when the combined count rates of all GBM detectors surpass this limit, data saturation occurs. We did not find saturation in the data during any of the 51 bursts in our chosen sample.


\section{Machine Learning Approach for Spectral Clustering} \label{appB} 

$K$-means clustering is a commonly used algorithm in data analysis that groups data points based on specific features, which in our case are the midpoints of time segments and corresponding $E_{\rm peak}$ values for each burst. Its primary objective is to cluster data points in a way that maximizes the similarity within clusters and minimizes the similarity between different clusters \citep{Lloyd_1982}.

In this study, Python programming language (version 3.6.9) and Scikit-learn (version 1.2.1) were used for $k$-means clustering. The algorithm requires a predetermined number of clusters ($k$). Firstly, it initializes centroids by randomly selecting $k$ samples from the data set. After initialization, the algorithm iterates between the remaining two steps: It first assigns each sample to the nearest centroid, then, creates new centroids by computing the mean value of all the samples assigned to each of the previous centroids. The algorithm repeats these last two steps until the squared difference between the old and new centroids is less than a predefined threshold, indicating that the centroids have become stable, and the clustering is complete.

For each burst, first, the data was scaled to prevent the effect of one variable from overriding the other since the ranges of time and $E_{\rm peak}$ axes vary significantly. Next, the $k$-means algorithm was implemented for all possible $k$ values ranging from one to ($N-1$), where \textit{N} is the number of time segments, and the corresponding inertia (i.e., the sum of squared distances of samples to their closest cluster center) for each $k$ was recorded. Note that the reciprocal of \ep errors was given as a weight for the inertia calculation. By definition of inertia, its value decreases as the number of clusters ($k$) increases. However, increasing $k$ does not significantly reduce inertia after a certain number of $k$, and the inertia vs. $k$ graph becomes flat after this point, encompassing a minimum of about 20$-$25\% of the high-\textit{k} values in our sample. After studying a sample of events, we found that this point corresponds to the optimal number of clusters for finding the largest spectral variations during each burst. Therefore, instead of heuristic techniques that are commonly used to find the optimal $k$ in $k$-means clustering, e.g., the ``elbow" method, we developed our method as follows: First, we calculated the average of inertia values that corresponds to the highest 25\% of the $k$ values in the inertia vs. $k$ graph. We then added 1\% of the maximum inertia value to the average inertia of the highest 25\% of the $k$ values. The nearest integer $k$ value corresponding to the resulting inertia is then used for the optimal number of clusters.

Finally, we note that in addition to $k$-means clustering we also tested other machine learning based clustering algorithms; namely, density-based spatial clustering, agglomerative clustering, and Gaussian mixture model. They all require a number of clusters to be specified and their results are consistent with those of $k$-means. We, therefore, consider $k$-means a robust method for our purpose.

\section{Continuum Model Comparisons} \label{appC} 

\begin{figure}[htbp]
    \centering
    \epsscale{1.15}
    \includegraphics[width = 14 cm, trim=60 280 45 265, clip]{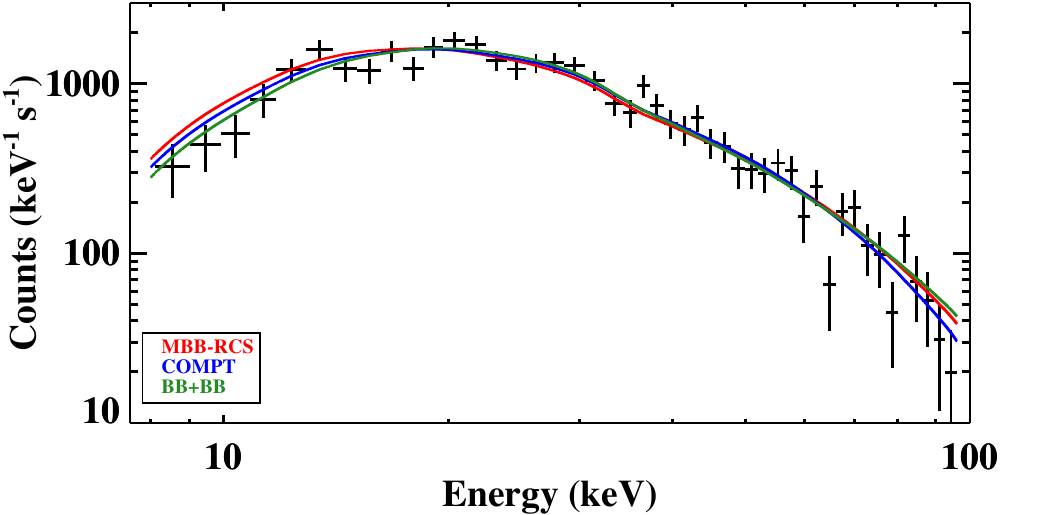}
       \caption{Observed count spectrum of a time segment from the burst detected at 488642074.718 (\textit{Fermi} MET) are shown in black crosses. The three solid trends are the best-fitting model curves: MBB-RCS in red, COMPT in blue, and BB+BB in green, all of which fit the spectrum equally well.
}
    \label{cnt_spec_all}
\end{figure}

To demonstrate how different photon models represent observed burst spectra, we present in Figure \ref{cnt_spec_all} the spectrum extracted from a time segment of the burst that occurred at 488642074.718 (\textit{Fermi} MET; 160626 13:54:30.722 UTC). All three models employed in this investigation yield equally well fit to this particular burst data in our energy passband: MBB-RCS with $kT$ of 9.46\,keV (red curve in Figure \ref{cnt_spec_all}), COMPT with $\Gamma$ = 0.56 and $E_{\rm peak}$ = 39.07\,keV (blue curve), and BB+BB with \lkt = 7.4 \& \hkt = 15.31\,keV (green curve).
Slight differences among the models are observed at the low-energy end and at the high-energy end, which are statistically indistinguishable.


\begin{figure}[htbp]
    \centering
    \epsscale{1.15}
    \includegraphics[width = 15 cm, trim=60 160 30 330, clip]{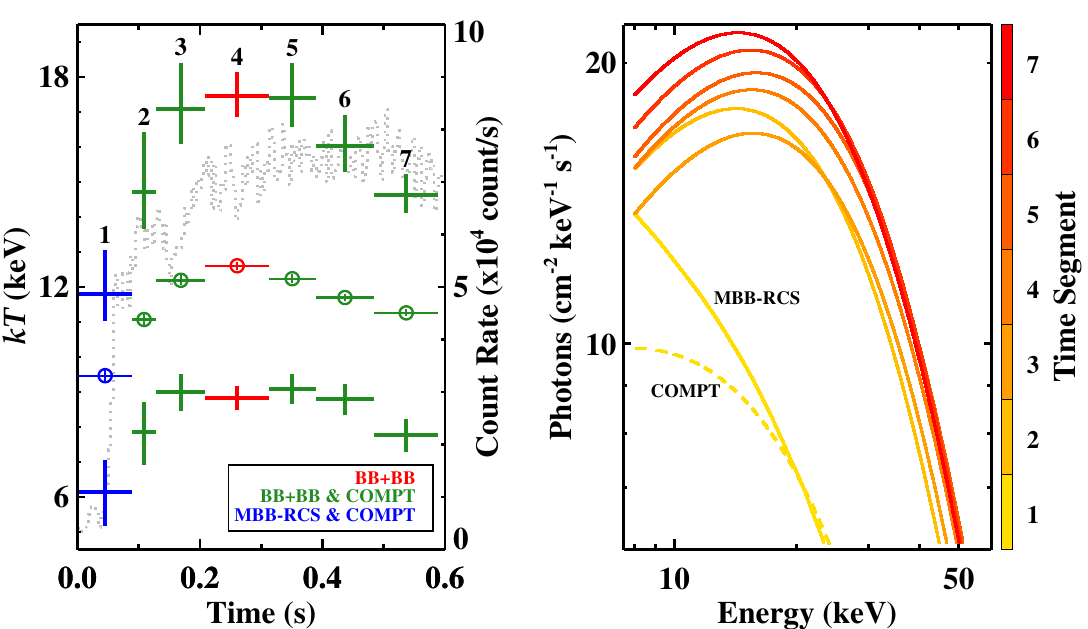}
       \caption{\textbf{[Left]} Blackbody temperature evolution of the burst detected at 652927551.870 (\textit{Fermi} MET; same event as Figures \ref{kmeans_plot} and \ref{thermal_plot}); this is the same plot as Figure \ref{thermal_plot} but zoomed into the first 0.6 s. The first seven time segments are numbered on top, and thick crosses show the \lkt and \hkt parameters of BB+BB while thin data points with circles show \mkt of MBB-RCS. The color code represents the preferred model(s). \textbf{[Right]} The photon flux distribution for each time segment is shown. The color code represents the time segments. For the first time segment, the photon flux shown with the yellow dashed line and solid line was obtained with COMPT (\ep = 37.85\,keV and $\Gamma$ = 0.54) and MBB-RCS, respectively. For the rest of the time segments, photon flux was obtained with BB+BB. 
}
    \label{cnt_spec_time}
\end{figure}

We also present the evolution of spectral parameters of another example burst detected at 652927551.870 (\textit{Fermi} MET; the same event also shown in Figure \ref{kmeans_plot} and \ref{thermal_plot}) in Figure \ref{cnt_spec_time}. This is an example event, throughout which a thermal model is preferred. In the left panel, we present the $kT$ parameters of the thermal models (i.e., BB+BB and MBB-RCS) for the first 7 time segments. In the right panel, we present the evolution of photon flux distribution with time (from yellow to red) using the BB+BB model for segments 2 through 7. Note that both COMPT and MBB-RCS model curves are displayed for the first time segment. Since photon flux distributions of segments 2 to 7 are similar at high energies (above 60\,keV), we zoomed into the energy range of $8 - 60$\,keV to clearly exhibit model differences at low energies. 


\bibliographystyle{aasjournal}
\bibliography{refs} 

\begin{thebibliography}{}
\expandafter\ifx\csname natexlab\endcsname\relax\def\natexlab#1{#1}\fi
\providecommand{\url}[1]{\href{#1}{#1}}
\providecommand{\dodoi}[1]{doi:~\href{http://doi.org/#1}{\nolinkurl{#1}}}
\providecommand{\doeprint}[1]{\href{http://ascl.net/#1}{\nolinkurl{http://ascl.net/#1}}}
\providecommand{\doarXiv}[1]{\href{https://arxiv.org/abs/#1}{\nolinkurl{https://arxiv.org/abs/#1}}}

\bibitem[{{Baring} \& {Harding}(2007)}]{BH-2007-ApandSS}
{Baring}, M.~G., \& {Harding}, A.~K. 2007, \apss, 308, 109, \dodoi{10.1007/s10509-007-9326-x}

\bibitem[{{Bochenek} {et~al.}(2020){Bochenek}, {Ravi}, {Belov}, {Hallinan}, {Kocz}, {Kulkarni}, \& {McKenna}}]{bochenek}
{Bochenek}, C.~D., {Ravi}, V., {Belov}, K.~V., {et~al.} 2020, \nat, 587, 59, \dodoi{10.1038/s41586-020-2872-x}

\bibitem[{{Cash}(1979)}]{Castor}
{Cash}, W. 1979, \apj, 228, 939, \dodoi{10.1086/156922}

\bibitem[{{CHIME/FRB Collaboration} {et~al.}(2020){CHIME/FRB Collaboration}, {Andersen}, {Bandura}, {Bhardwaj}, {Bij}, {Boyce}, {Boyle}, {Brar}, {Cassanelli}, {Chawla}, {Chen}, {Cliche}, {Cook}, {Cubranic}, {Curtin}, {Denman}, {Dobbs}, {Dong}, {Fandino}, {Fonseca}, {Gaensler}, {Giri}, {Good}, {Halpern}, {Hill}, {Hinshaw}, {H{\"o}fer}, {Josephy}, {Kania}, {Kaspi}, {Landecker}, {Leung}, {Li}, {Lin}, {Masui}, {McKinven}, {Mena-Parra}, {Merryfield}, {Meyers}, {Michilli}, {Milutinovic}, {Mirhosseini}, {M{\"u}nchmeyer}, {Naidu}, {Newburgh}, {Ng}, {Patel}, {Pen}, {Pinsonneault-Marotte}, {Pleunis}, {Quine}, {Rafiei-Ravandi}, {Rahman}, {Ransom}, {Renard}, {Sanghavi}, {Scholz}, {Shaw}, {Shin}, {Siegel}, {Singh}, {Smegal}, {Smith}, {Stairs}, {Tan}, {Tendulkar}, {Tretyakov}, {Vanderlinde}, {Wang}, {Wulf}, \& {Zwaniga}}]{chime20}
{CHIME/FRB Collaboration}, {Andersen}, B.~C., {Bandura}, K.~M., {et~al.} 2020, \nat, 587, 54, \dodoi{10.1038/s41586-020-2863-y}

\bibitem[{{den Hartog} {et~al.}(2008){den Hartog}, {Kuiper}, {Hermsen}, {Kaspi}, {Dib}, {Kn{\"o}dlseder}, \& {Gavriil}}]{denHartog-2008-AandA}
{den Hartog}, P.~R., {Kuiper}, L., {Hermsen}, W., {et~al.} 2008, \aap, 489, 245, \dodoi{10.1051/0004-6361:200809390}

\bibitem[{{Fern{\'a}ndez} \& {Thompson}(2007)}]{Fernandez-2007-ApJ}
{Fern{\'a}ndez}, R., \& {Thompson}, C. 2007, \apj, 660, 615, \dodoi{10.1086/511810}

\bibitem[{{G{\"o}tz} {et~al.}(2006){G{\"o}tz}, {Mereghetti}, {Tiengo}, \& {Esposito}}]{Goetz-2006-AandA}
{G{\"o}tz}, D., {Mereghetti}, S., {Tiengo}, A., \& {Esposito}, P. 2006, \aap, 449, L31, \dodoi{10.1051/0004-6361:20064870}

\bibitem[{{G{\"o}{\v{g}}{\"u}{\c{s}}} {et~al.}(2001){G{\"o}{\v{g}}{\"u}{\c{s}}}, {Kouveliotou}, {Woods}, {Thompson}, {Duncan}, \& {Briggs}}]{EG01}
{G{\"o}{\v{g}}{\"u}{\c{s}}}, E., {Kouveliotou}, C., {Woods}, P.~M., {et~al.} 2001, \apj, 558, 228, \dodoi{10.1086/322463}

\bibitem[{{Hu} {et~al.}(2022){Hu}, {Baring}, {Barchas}, \& {Younes}}]{Hu-2022-ApJ}
{Hu}, K., {Baring}, M.~G., {Barchas}, J.~A., \& {Younes}, G. 2022, \apj, 928, 82, \dodoi{10.3847/1538-4357/ac4ae8}

\bibitem[{{Hurley} {et~al.}(1999){Hurley}, {Cline}, {Mazets}, {Barthelmy}, {Butterworth}, {Marshall}, {Palmer}, {Aptekar}, {Golenetskii}, {Il'Inskii}, {Frederiks}, {McTiernan}, {Gold}, \& {Trombka}}]{Hurley99}
{Hurley}, K., {Cline}, T., {Mazets}, E., {et~al.} 1999, \nat, 397, 41, \dodoi{10.1038/16199}

\bibitem[{{Ibrahim} {et~al.}(2001){Ibrahim}, {Strohmayer}, {Woods}, {Kouveliotou}, {Thompson}, {Duncan}, {Dieters}, {Swank}, {van Paradijs}, \& {Finger}}]{Alaa01}
{Ibrahim}, A.~I., {Strohmayer}, T.~E., {Woods}, P.~M., {et~al.} 2001, \apj, 558, 237, \dodoi{10.1086/322248}

\bibitem[{{Israel} {et~al.}(2008){Israel}, {Romano}, {Mangano}, {Dall'Osso}, {Chincarini}, {Stella}, {Campana}, {Belloni}, {Tagliaferri}, {Blustin}, {Sakamoto}, {Hurley}, {Zane}, {Moretti}, {Palmer}, {Guidorzi}, {Burrows}, {Gehrels}, \& {Krimm}}]{Israel08}
{Israel}, G.~L., {Romano}, P., {Mangano}, V., {et~al.} 2008, \apj, 685, 1114, \dodoi{10.1086/590486}

\bibitem[{{Israel} {et~al.}(2016){Israel}, {Esposito}, {Rea}, {Coti Zelati}, {Tiengo}, {Campana}, {Mereghetti}, {Rodriguez Castillo}, {G{\"o}tz}, {Burgay}, {Possenti}, {Zane}, {Turolla}, {Perna}, {Cannizzaro}, \& {Pons}}]{Israel06}
{Israel}, G.~L., {Esposito}, P., {Rea}, N., {et~al.} 2016, \mnras, 457, 3448, \dodoi{10.1093/mnras/stw008}

\bibitem[{{Kaastra}(2017)}]{Kaastra2017}
{Kaastra}, J.~S. 2017, \aap, 605, A51, \dodoi{10.1051/0004-6361/201629319}

\bibitem[{{Kaneko} {et~al.}(2021){Kaneko}, {G{\"o}{\u{g}}{\"u}{\c{s}}}, {Baring}, {Kouveliotou}, {Lin}, {Roberts}, {van der Horst}, {Younes}, {Keskin}, \& {{\c{C}}oban}}]{YK21}
{Kaneko}, Y., {G{\"o}{\u{g}}{\"u}{\c{s}}}, E., {Baring}, M.~G., {et~al.} 2021, \apjl, 916, L7, \dodoi{10.3847/2041-8213/ac0fe7}

\bibitem[{Kass \& Raftery(1995)}]{Bayes95}
Kass, R.~E., \& Raftery, A.~E. 1995, Journal of the American Statistical Association, 90, 773, \dodoi{10.1080/01621459.1995.10476572}

\bibitem[{{Kouveliotou} {et~al.}(2001){Kouveliotou}, {Tennant}, {Woods}, {Weisskopf}, {Hurley}, {Fender}, {Garrington}, {Patel}, \& {G{\"o}{\v{g}}{\"u}{\c{s}}}}]{CK01}
{Kouveliotou}, C., {Tennant}, A., {Woods}, P.~M., {et~al.} 2001, \apjl, 558, L47, \dodoi{10.1086/323496}

\bibitem[{{Kozlova} {et~al.}(2016){Kozlova}, {Israel}, {Svinkin}, {Frederiks}, {Pal'shin}, {Tsvetkova}, {Hurley}, {Goldsten}, {Golovin}, {Mitrofanov}, \& {Zhang}}]{Kozlova16}
{Kozlova}, A.~V., {Israel}, G.~L., {Svinkin}, D.~S., {et~al.} 2016, \mnras, 460, 2008, \dodoi{10.1093/mnras/stw1109}

\bibitem[{{Kuiper} {et~al.}(2004){Kuiper}, {Hermsen}, \& {Mendez}}]{Kuiper-2004-ApJ}
{Kuiper}, L., {Hermsen}, W., \& {Mendez}, M. 2004, \apj, 613, 1173, \dodoi{10.1086/423129}

\bibitem[{{Lander}(2023)}]{Lander23}
{Lander}, S.~K. 2023, \apjl, 947, L16, \dodoi{10.3847/2041-8213/acca1f}

\bibitem[{{Liddle}(2007)}]{Liddle07}
{Liddle}, A.~R. 2007, \mnras, 377, L74, \dodoi{10.1111/j.1745-3933.2007.00306.x}

\bibitem[{{Lin} {et~al.}(2020{\natexlab{a}}){Lin}, {G{\"o}{\u{g}}{\"u}{\c{s}}}, {Roberts}, {Kouveliotou}, {Kaneko}, {van der Horst}, \& {Younes}}]{Lin2016}
{Lin}, L., {G{\"o}{\u{g}}{\"u}{\c{s}}}, E., {Roberts}, O.~J., {et~al.} 2020{\natexlab{a}}, \apj, 893, 156, \dodoi{10.3847/1538-4357/ab818f}

\bibitem[{{Lin} {et~al.}(2020{\natexlab{b}}){Lin}, {G{\"o}{\u{g}}{\"u}{\textcommabelow s}}, {Roberts}, {Baring}, {Kouveliotou}, {Kaneko}, {van der Horst}, \& {Younes}}]{Lin2020}
{Lin}, L., {G{\"o}{\u{g}}{\"u}{\textcommabelow s}}, E., {Roberts}, O.~J., {et~al.} 2020{\natexlab{b}}, \apjl, 902, L43, \dodoi{10.3847/2041-8213/abbefe}

\bibitem[{{Lin} {et~al.}(2013){Lin}, {G{\"o}{\v{g}}{\"u}{\c{s}}}, {Kaneko}, \& {Kouveliotou}}]{Lin2013}
{Lin}, L., {G{\"o}{\v{g}}{\"u}{\c{s}}}, E., {Kaneko}, Y., \& {Kouveliotou}, C. 2013, \apj, 778, 105, \dodoi{10.1088/0004-637X/778/2/105}

\bibitem[{{Lin} {et~al.}(2011){Lin}, {Kouveliotou}, {Baring}, {van der Horst}, {Guiriec}, {Woods}, {G{\"o}{\v{g}}{\"u}{\c{s}}}, {Kaneko}, {Scargle}, {Granot}, {Preece}, {von Kienlin}, {Chaplin}, {Watts}, {Wijers}, {Zhang}, {Bhat}, {Finger}, {Gehrels}, {Harding}, {Kaper}, {Kaspi}, {Mcenery}, {Meegan}, {Paciesas}, {Pe'er}, {Ramirez-Ruiz}, {van der Klis}, {Wachter}, \& {Wilson-Hodge}}]{Lin2011}
{Lin}, L., {Kouveliotou}, C., {Baring}, M.~G., {et~al.} 2011, \apj, 739, 87, \dodoi{10.1088/0004-637X/739/2/87}

\bibitem[{{Lin} {et~al.}(2012){Lin}, {G{\"o}{\v{g}}{\"u}{\c{s}}}, {Baring}, {Granot}, {Kouveliotou}, {Kaneko}, {van der Horst}, {Gruber}, {von Kienlin}, {Younes}, {Watts}, \& {Gehrels}}]{Lin12}
{Lin}, L., {G{\"o}{\v{g}}{\"u}{\c{s}}}, E., {Baring}, M.~G., {et~al.} 2012, \apj, 756, 54, \dodoi{10.1088/0004-637X/756/1/54}

\bibitem[{Lloyd(1982)}]{Lloyd_1982}
Lloyd, S. 1982, IEEE Transactions on Information Theory, 28, 129–137, \dodoi{10.1109/tit.1982.1056489}

\bibitem[{Lyubarsky(2002)}]{Lyubarsky_2002}
Lyubarsky, Y.~E. 2002, Monthly Notices of the Royal Astronomical Society, 332, 199–204, \dodoi{10.1046/j.1365-8711.2002.05290.x}

\bibitem[{{Lyutikov}(2003)}]{lyu03}
{Lyutikov}, M. 2003, \mnras, 346, 540, \dodoi{10.1046/j.1365-2966.2003.07110.x}

\bibitem[{{Meegan} {et~al.}(2009){Meegan}, {Lichti}, {Bhat}, {Bissaldi}, {Briggs}, {Connaughton}, {Diehl}, {Fishman}, {Greiner}, {Hoover}, {van der Horst}, {von Kienlin}, {Kippen}, {Kouveliotou}, {McBreen}, {Paciesas}, {Preece}, {Steinle}, {Wallace}, {Wilson}, \& {Wilson-Hodge}}]{Meegan2009}
{Meegan}, C., {Lichti}, G., {Bhat}, P.~N., {et~al.} 2009, \apj, 702, 791, \dodoi{10.1088/0004-637X/702/1/791}

\bibitem[{{Mereghetti} {et~al.}(2020){Mereghetti}, {Savchenko}, {Ferrigno}, {G{\"o}tz}, {Rigoselli}, {Tiengo}, {Bazzano}, {Bozzo}, {Coleiro}, {Courvoisier}, {Doyle}, {Goldwurm}, {Hanlon}, {Jourdain}, {von Kienlin}, {Lutovinov}, {Martin-Carrillo}, {Molkov}, {Natalucci}, {Onori}, {Panessa}, {Rodi}, {Rodriguez}, {S{\'a}nchez-Fern{\'a}ndez}, {Sunyaev}, \& {Ubertini}}]{mereghetti20}
{Mereghetti}, S., {Savchenko}, V., {Ferrigno}, C., {et~al.} 2020, \apjl, 898, L29, \dodoi{10.3847/2041-8213/aba2cf}

\bibitem[{{Palmer} {et~al.}(2005){Palmer}, {Barthelmy}, {Gehrels}, {Kippen}, {Cayton}, {Kouveliotou}, {Eichler}, {Wijers}, {Woods}, {Granot}, {Lyubarsky}, {Ramirez-Ruiz}, {Barbier}, {Chester}, {Cummings}, {Fenimore}, {Finger}, {Gaensler}, {Hullinger}, {Krimm}, {Markwardt}, {Nousek}, {Parsons}, {Patel}, {Sakamoto}, {Sato}, {Suzuki}, \& {Tueller}}]{Palmer05}
{Palmer}, D.~M., {Barthelmy}, S., {Gehrels}, N., {et~al.} 2005, \nat, 434, 1107, \dodoi{10.1038/nature03525}

\bibitem[{Pedregosa {et~al.}(2011)Pedregosa, Varoquaux, Gramfort, Michel, Thirion, Grisel, Blondel, Prettenhofer, Weiss, Dubourg, Vanderplas, Passos, Cournapeau, Brucher, Perrot, \& {{\'E}}douard Duchesnay}]{scikit-learn}
Pedregosa, F., Varoquaux, G., Gramfort, A., {et~al.} 2011, Journal of Machine Learning Research, 12, 2825.
\newblock \url{http://jmlr.org/papers/v12/pedregosa11a.html}

\bibitem[{{Petroff} {et~al.}(2019){Petroff}, {Hessels}, \& {Lorimer}}]{petroff}
{Petroff}, E., {Hessels}, J.~W.~T., \& {Lorimer}, D.~R. 2019, \aapr, 27, 4, \dodoi{10.1007/s00159-019-0116-6}

\bibitem[{{Petroff} {et~al.}(2021){Petroff}, {Hessels}, \& {Lorimer}}]{petroff21}
---. 2021, arXiv e-prints, arXiv:2107.10113.
\newblock \doarXiv{2107.10113}

\bibitem[{{Scargle} {et~al.}(2013){Scargle}, {Norris}, {Jackson}, \& {Chiang}}]{scargle2013}
{Scargle}, J.~D., {Norris}, J.~P., {Jackson}, B., \& {Chiang}, J. 2013, \apj, 764, 167, \dodoi{10.1088/0004-637X/764/2/167}

\bibitem[{Schwarz(1978)}]{Schwarz1978}
Schwarz, G. 1978, The Annals of Statistics, 6, 461

\bibitem[{{Strohmayer} {et~al.}(1998){Strohmayer}, {Zhang}, {Swank}, {White}, \& {Lapidus}}]{Stroh98}
{Strohmayer}, T.~E., {Zhang}, W., {Swank}, J.~H., {White}, N.~E., \& {Lapidus}, I. 1998, \apjl, 498, L135, \dodoi{10.1086/311322}

\bibitem[{{Thompson} \& {Duncan}(1995)}]{TD95}
{Thompson}, C., \& {Duncan}, R.~C. 1995, \mnras, 275, 255, \dodoi{10.1093/mnras/275.2.255}

\bibitem[{{Thompson} \& {Duncan}(1996)}]{TD-1996-ApJ}
---. 1996, \apj, 473, 322, \dodoi{10.1086/178147}

\bibitem[{{Thompson} \& {Duncan}(2001)}]{TD01}
---. 2001, \apj, 561, 980, \dodoi{10.1086/323256}

\bibitem[{{van der Horst} {et~al.}(2012){van der Horst}, {Kouveliotou}, {Gorgone}, {Kaneko}, {Baring}, {Guiriec}, {G{\"o}{\v{g}}{\"u}{\c{s}}}, {Granot}, {Watts}, {Lin}, {Bhat}, {Bissaldi}, {Chaplin}, {Finger}, {Gehrels}, {Gibby}, {Giles}, {Goldstein}, {Gruber}, {Harding}, {Kaper}, {von Kienlin}, {van der Klis}, {McBreen}, {Mcenery}, {Meegan}, {Paciesas}, {Pe'er}, {Preece}, {Ramirez-Ruiz}, {Rau}, {Wachter}, {Wilson-Hodge}, {Woods}, \& {Wijers}}]{von12}
{van der Horst}, A.~J., {Kouveliotou}, C., {Gorgone}, N.~M., {et~al.} 2012, \apj, 749, 122, \dodoi{10.1088/0004-637X/749/2/122}

\bibitem[{{Wadiasingh} {et~al.}(2018){Wadiasingh}, {Baring}, {Gonthier}, \& {Harding}}]{Wadiasingh-2018-ApJ}
{Wadiasingh}, Z., {Baring}, M.~G., {Gonthier}, P.~L., \& {Harding}, A.~K. 2018, \apj, 854, 98, \dodoi{10.3847/1538-4357/aaa460}

\bibitem[{{Yamasaki} {et~al.}(2020){Yamasaki}, {Lyubarsky}, {Granot}, \& {G{\"o}{\u{g}}{\"u}{\c{s}}}}]{Yamasaki20}
{Yamasaki}, S., {Lyubarsky}, Y., {Granot}, J., \& {G{\"o}{\u{g}}{\"u}{\c{s}}}, E. 2020, \mnras, 498, 484, \dodoi{10.1093/mnras/staa2223}

\bibitem[{{Younes} {et~al.}(2014){Younes}, {Kouveliotou}, {van der Horst}, {Baring}, {Granot}, {Watts}, {Bhat}, {Collazzi}, {Gehrels}, {Gorgone}, {G{\"o}{\u{g}}{\"u}{\c{s}}}, {Gruber}, {Grunblatt}, {Huppenkothen}, {Kaneko}, {von Kienlin}, {van der Klis}, {Lin}, {Mcenery}, {van Putten}, \& {Wijers}}]{Younes14}
{Younes}, G., {Kouveliotou}, C., {van der Horst}, A.~J., {et~al.} 2014, \apj, 785, 52, \dodoi{10.1088/0004-637X/785/1/52}

\end{thebibliography}

\end{document}